\documentclass[prd,twocolumn,showpacs]{revtex4-1}
\usepackage[utf8]{inputenc}
\usepackage{amsmath}
\usepackage{latexsym}
\usepackage{amsfonts}
\usepackage{graphicx}
\usepackage{mathrsfs}
\usepackage{CJK}
\usepackage{longtable}
\usepackage[normalem]{ulem}

\usepackage{hyperref}
\hypersetup{
  colorlinks = true,
  urlcolor = blue,
  linkcolor = blue,
  citecolor = green,
  filecolor = magenta,
}

\newcommand{\ud}{\mathrm{d}}

\begin{document}
\begin{CJK*}{GB}{song}

\title{Gravitational Wave Interference via Gravitational Lensing: Measurements of Luminosity Distance, Lens Mass, and Cosmological Parameters}
\author{Shaoqi Hou}
\affiliation{School of Physics and Technology, Wuhan University, Wuhan, Hubei 430072, China}
\author{Xi-Long Fan}
\email{xilong.fan@whu.edu.cn}
\affiliation{School of Physics and Technology, Wuhan University, Wuhan, Hubei 430072, China}
\author{Kai Liao}
\affiliation{School of Physics and Technology, Wuhan University, Wuhan, Hubei 430072, China}
\author{Zong-Hong Zhu}
\email{zhuzh@whu.edu.cn}
\affiliation{School of Physics and Technology, Wuhan University, Wuhan, Hubei 430072, China}
\affiliation{Department of Astronomy, Beijing Normal University, Beijing 100875,  China}
\date{\today}

\begin{abstract}
The gravitational lensing of gravitational waves might cause beat patterns detectable by interferometers. 
The feature of this kind of signal is the existence of the beat pattern in the early inspiral phase, followed by a seemingly randomly changing profile. 
After the strain peaks for the first time, the signal takes the usual waveform and the strain peaks for the second time. 
Once this signal is detected, the actual magnification factors can be obtained, so the true luminosity distance of the binary system is known. 
If the lens can be described by a point mass or a singular isothermal sphere, the functional forms of the time delay and the magnification factors are simple enough, so we can infer the mass of the lens or the cosmological parameters.
\end{abstract}

\maketitle

\section{Introduction} 
\label{sec-int}

The existence of the gravitational wave (GW) has been confirmed by the 12 GW events observed by the LIGO and Virgo (LIGO/Virgo) collaborations \cite{Abbott:2016blz,*Abbott:2016nmj,*Abbott:2017vtc,*Abbott:2017oio,*TheLIGOScientific:2017qsa,*Abbott:2017gyy,*LIGOScientific:2018mvr,*Abbott:2020uma}.
Although modified theories of gravity also predict GWs \cite{dePaula:2004bc,*Nishizawa:2009bf,*Will:2014kxa,*Hou:2017bqj,*Hou:2018djz,*Gong:2018ybk,*Gong:2018cgj,*Hou:2018mey,*Soudi:2018dhv,*Hohmann:2019nat,*Liu:2019cxm}, we focus on general relativity.
Like the electromagnetic wave, GWs propagate at the speed of light $c$ and polarize in the direction perpendicular to their propagation directions \cite{Misner:1974qy}.
Because of the equivalence principle, the trajectory of the GW bends when there is a gravitational potential near its way.
This leads to the gravitational lensing effect \cite{gravlens1992,Lawrence1971nc,*Lawrence:1971hx}.
The gravitational lensing could cause the magnification in the GW amplitude, making it easier to detect.
However, this also leads to the underestimate of the luminosity distance of the GW source \cite{Ng:2017yiu,Dai:2016igl}.

GWs produced by the binary systems have much longer wavelengths than the (visible) light, so the wave optics effects \cite{gravlens1992,Nakamura:1997sw,Nakamura1999wo,Takahashi:2003ix,Takahashi:2016jom} would be strong if the lens is not massive enough. 
For GWs detectable by the ground-based detectors (1 - $10^4$ Hz), the geometric optics works if the mass of the lens is larger than about $10^4M_\odot$, while for GWs with the frequencies in the LISA band ($10^{-4}$ - $10^{-1}$ Hz), the lens mass  should be at least $10^8M_\odot$  \cite{ArnaudVarvella:2003va,Meena:2019ate}.
Of course, if DECi-hertz Interferometer Gravitational wave Observatory (DECIGO) \cite{Kawamura:2011zz} is used to look for lensed GWs, the lower bound on the lens mass should be somewhere in between.
Despite some interesting phenomena (e.g., diffraction) due to the wave optics \cite{Nakamura1999wo,Liao:2019aqq}, we would like to concentrate on the geometric optics. 
In this case, the lensed GWs travel in distinct paths and arrive at the Earth at different times. 

In this work, we will consider the strong lensing effect of the GW.
More specifically, the time delay $\Delta t$ between lensed GWs should be on the order of merely a few seconds if they are to be detected by the ground-based interferometers, while if the space-borne detectors are used, $\Delta t$ can be as long as a few months.
In this case, these detectors would observe the lensed GWs simultaneously.
Since the binary system produces nearly monochromatic GWs and the GWs emitted in distinct directions carry definite phase differences from each other \cite{Hou:2019wdg}, a beat pattern appears in the time domain.
Gradually, the beat pattern fades, since the frequency difference grows larger compared to their average.
After the merger is observed along the earlier GW, the beat pattern quickly diminishes, and the usual waveform emerges.
As long as  such a kind of behavior is observed, one infers that the GWs were \textit{probably} lensed \cite{Nakamura1999wo}.
This scenario allows us to infer the magnification factors of the amplitudes, and then the absolute magnification factors can be determined,  which gives the actual luminosity distance of the GW source. 
One can also directly measure $\Delta t$ very accurately from the observed waveform without the aid of the electromagnetic counterparts.
With this information and assuming suitable lens models, the redshifted lens mass or the cosmological parameters can be determined. 

To form the beat pattern, the constraint on $\Delta t$ for the ground-based detectors is highly tight, so the probability of using the ground-based detectors to find beats is tiny. 
However, the requirement on $\Delta t$ for the space-borne detectors is very relaxed, which implies that it is more plausible to use the detectors such as LISA or DECIGO to detect the beat pattern.
In spite of the low probability associated with the ground-based detectors, we will start with the discussion on the beats detected by these detectors. 
This is because during the observation period the motion of the Earth can be ignored, which makes the analysis simpler. 
The analysis of the beats observed by the space-borne detectors is more complicated as the detectors are moving in space during the observation, and their motion constantly changes the antenna pattern functions \cite{Schilling:1997id,*Giampieri:1997kv,*Liang:2019pry}. 
But if the beat period is in the range of a few hours to a few minutes, it is a good approximation to ignore the motion of the space-borne detectors. 
So once we understand how the ground-based detectors detect the beat pattern and make use of it, we can easily generalize that to the space-borne detectors.

Moreover, in the discussion on the beat pattern observed by the ground-based detectors, we use the simple lens model of a point mass to illustrate the formation of the beats, the conditions, and the applications.
In contrast, for the space-borne detectors, the lens  is chosen to be a singular isothermal sphere  (SIS) \cite{gravlens1992}.
The stars in galaxies may cause microlensing that also leads to the modulation in the GW amplitude.
By Ref.~\cite{Diego:2019lcd}, the effect of the microlensing is prominent when the magnification of the strong lensing is about a few hundreds to a few thousands, that is, when the saturation regime \footnote{According to Ref.~\cite{Diego:2019lcd}, in the saturation regime, the caustics corresponding to the microlenses contained in the galaxies overlap in the source plane. 
So  GWs emitted by sources near or moving across this regime will be microlensed. 
Refer to Ref.~\cite{Diego:2017drh}, too.} is reached. 
In the following scenarios to be discussed, the saturation is never reached, so the microlensing effect is negligible.
Even if the magnification of the strong lensing is very large in some situations, the microlensing effect might also be less significant, because the beat pattern considered in our work exists in the (early) inspiral phase, and the microlensing induces modulation in higher frequency range ($\sim$a few hundred hertz) \cite{Diego:2019lcd}.
Therefore, in our discussion, we will ignore the microlensing effect.

GW sources can be used as the standard sirens to measure their luminosity distances \cite{Schutz:1986gp}.
However, the gravitational lensing, in particular, the weak lensing, makes  the measurement less accurate with the uncertainty around a few percent \cite{Holz:2005df}. 
This can be partially mitigated by utilizing the shear and flexion maps to infer the convergence and thus the magnification \cite{Shapiro:2009sr,Hilbert:2010am}.
The uncertainty in the luminosity distance can be reduced by 50\%.
The remaining uncertainty will inevitably affect our analysis, which should be considered in more realistic analysis. 
In this work, we neglect the effect of the weak lensing. 


In Ref.~\cite{Yamamoto:2005ea}, the interference between lensed GWs was also investigated within the geometric optics regime.
The author did not use this interference to determine the luminosity distance or  the lens mass or infer cosmological implications.
Gravitational lensing has a wide range of applications, for example, detecting dark matter \cite{Cutler:2009qv,Camera:2013xfa,Congedo:2018wfn,Jung:2017flg}, constraining the speed of light \cite{Fan:2016swi,Collett:2016dey}, determining the cosmological constant \cite{Sereno:2010dr,Sereno:2011ty,Liao:2017ioi}, and examining the wave nature of GWs \cite{Dai:2018enj,Liao:2019aqq,Sun:2019ztn}.
Although no gravitational lensing signals have been detected in the observed GW events, the advent of more sensitive GW detectors might make it possible soon \cite{Hannuksela:2019kle}.

In the following, we start with a brief review on the gravitational lensing of GWs in Sec.~\ref{sec-glgw}.
Then, we discuss the formation of the beat pattern observable by the ground-based interferometers in Sec.~\ref{sec-iog}.
There, Sec.~\ref{sec-beat-g} discusses generally the formation and the features of the beat pattern; Sec.~\ref{sec-snr-g} is devoted to the signal-to-noise ratio (SNR) of such a kind of signal; and finally, Sec.~\ref{sec-b-app} focuses on the application of the beat pattern to determine the redshifted lens mass. 
In Sec.~\ref{sec-ios}, we generalize the previous analysis to the space-borne detectors, in particular, LISA.
In Sec.~\ref{sec-beat-s}, the formation and the SNR of the beat detected by LISA are presented, and in Sec.~\ref{sec-be-s-app}, the application of the beat pattern to constrain cosmological parameters is speculated. 
Of course, in both Secs.~\ref{sec-iog} and \ref{sec-ios}, how to determine the actual luminosity distance of the GW source is discussed. 
Finally, Sec.~\ref{sec-con} summarizes this work.
In this work, the geometric units ($G=c=1$) are used.

\section{Gravitational lensing of gravitational waves} 
\label{sec-glgw}

In the geometric optics limit, GWs propagating in a generic, curved background interact with the background and experience three effects. 
First, the GW is described by  gravitons, traveling in  null geodesics.
Second, the number of gravitons is conserved  along the trajectory, and the polarization plane rotates as the trajectory bends \cite{Hou:2019wdg}.
Since the deflection angle is very small,  this rotation is ignored in this work.
Third, the gravitational Faraday rotation occurs at the higher orders, and so will also be ignored \cite{Piran1985nf,*Piran:1985dk,*Wang:1991nf}.

\begin{figure}
  \centering
  \includegraphics[width=0.4\textwidth]{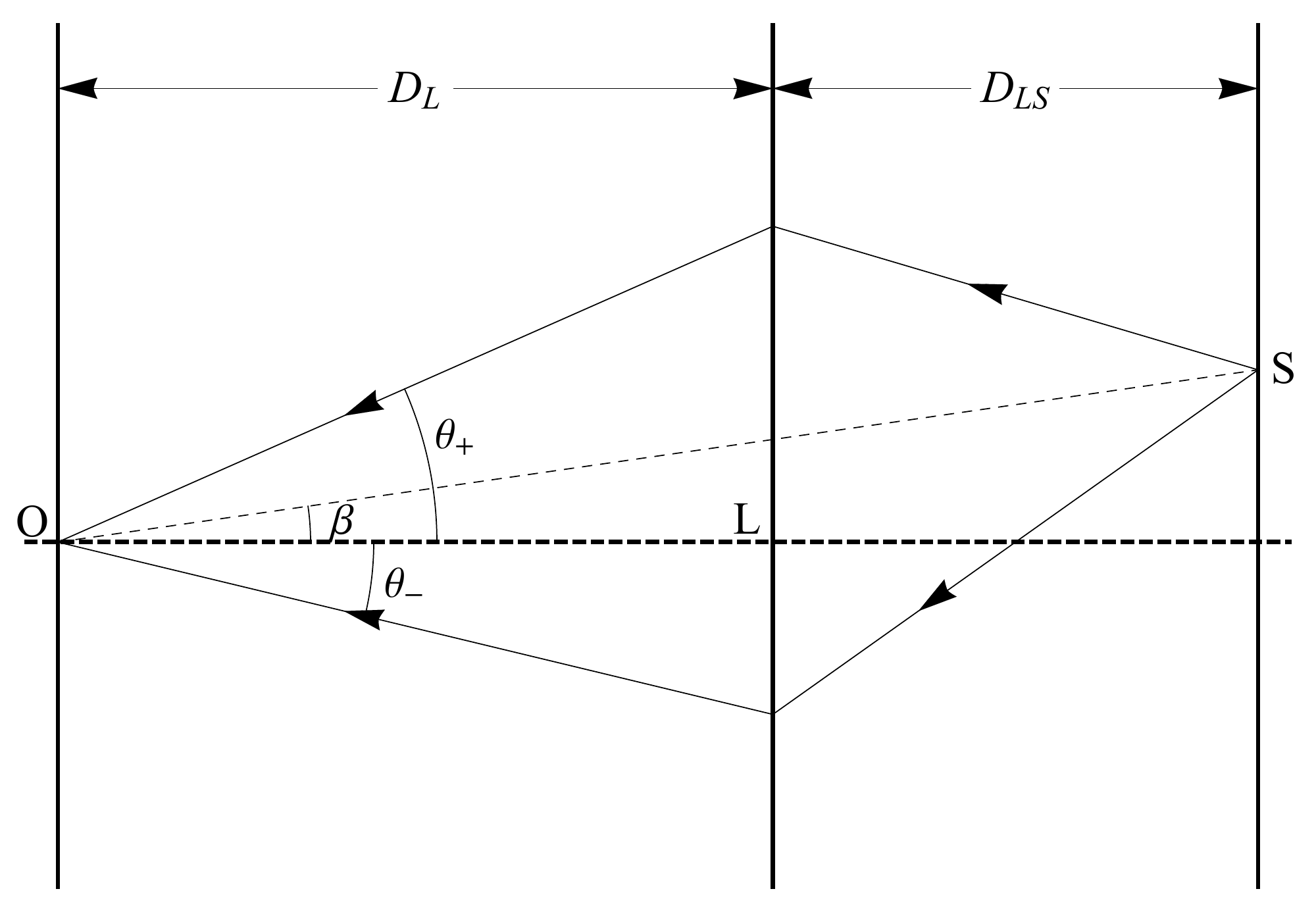}
  \caption{Geometry of a Schwarzschild lens.}\label{fig-geogl}
\end{figure}

Figure~\ref{fig-geogl} shows a typical geometry of a lens.
The lens is labeled by L, and two GW rays (1 and 2) pass by it in two trajectories. 
The angles $\theta_\pm$ are between the lensed rays and the optical axis OL, with O labeling the observer.
There are many different lens models. 
In the simplest case, the lens is a point mass $M$.
Then, one has \cite{gravlens1992}
\begin{equation}\label{eq-dan}
\theta_\pm=\frac{1}{2}\left(\beta\pm\sqrt{\beta^2+4\theta_\text{E}^2}\right),
\end{equation}
where $\beta$ is the misalignment angle between the optical axis and the direction from the observer to the source of the GW, and $\theta_\text{E}=\sqrt{4M\frac{D_\text{LS}}{D_\text{S}D_\text{L}}}$ is the Einstein angle.
$D_\text{L}=D(z_\text{L})$ and $D_\text{S}=D(z_\text{S})$ are the angular diameter distances $D(z)$ at redshifts $z_\text{L}$ and $z_\text{S}$, respectively.
$D_\text{LS}$ is the angular diameter distance between the source and the lens.
Since the probabilities for lensed GWs from the neutron star-neutron star mergers and the black hole-black hole mergers peak at redshifts about 2 and 4  \cite{Yang:2019jhw}, respectively, in this work, we are considering the sources at about $z=2$ as examples.
Because of the focusing effect of the lens, the amplitudes of the two rays get magnified by factors of
\begin{equation}\label{eq-mf}
{\mu_\pm}=\frac{|\theta_\pm|}{\sqrt{|\theta_+^2-\theta_-^2|}},
\end{equation}
respectively.
Finally, there is a time delay between the two rays, given by
\begin{equation}\label{eq-tdv}
  \Delta t  =4M(1+z_\text{L})\left(\frac{\theta_+^2-\theta_-^2}{2\theta_\text{E}^2}+\ln\frac{\theta_+}{-\theta_-}\right).
\end{equation}

A more realistic lens model is the SIS, which is suitable for the early-type galaxies, and has a major contribution to the strong lensing probability \cite{Turner:1984ch,Moeller:2006cu}.
In this model, 
\begin{equation}
  \label{eq-ang-sis}
 \theta_\pm=\beta\pm\theta_\text{E},
\end{equation}
where the Einstein angle is $\theta_\text{E}=4\pi\sigma_v^2D_\text{LS}/D_\text{S}$ with $\sigma_v$ being the line-of-sight velocity dispersion of the stars in the galaxy. 
The magnification factors are 
\begin{equation}\label{eq-mu-sis}
  \mu_\pm=\sqrt{\left|\frac{\theta_\pm/\theta_\text{E}}{|\theta_\pm/\theta_\text{E}|-1}\right|},
\end{equation}
and the time delay is 
\begin{equation}\label{eq-tdv-sis}
  \Delta t  =16\pi^2\sigma_v^4(1+z_\text{L})\frac{D_\text{L}D_\text{LS}}{D_\text{S}}\frac{\theta_++\theta_-}{\theta_\text{E}}.
\end{equation}

In both models,  the amplitudes of GWs get amplified by factors of ${\mu_\pm}$ after passing the lens.
This would cause the underestimation of the luminosity distance of the source if the lensing effect were ignored. 
The time delays of many lensing systems are much longer than the observation periods of LIGO/Virgo, but usually shorter than that of LISA, depending on the mass function \cite{Oguri:2018muv}.
The particular scenario to be considered below provides the possibility to take account of the effect of the lensing, and the true luminosity distance can be inferred, because the GW rays reach the detector simultaneously in certain time windows.
Moreover, the luminosity distance, the lens mass and the cosmological parameters could also be obtained in the following scenario.

\section{Interference observed by the ground-based detectors}  
\label{sec-iog}

\subsection{Beat pattern}
\label{sec-beat-g}


If there is a time window when the GW rays 1 and 2 reach the detector simultaneously, the strain is $h(t)=h_1(t)+h_2(t)$, which can be schematically expressed as
\begin{equation}\label{eq-s-12}
\begin{split}
  h=&{\mu_+}\left[A^+\cos(\omega_1t+\phi_1)+A^\times\sin\left(\omega_1t+\phi_1\right)\right]\\
  &+{\mu_-}\left[A^+\cos(\omega_2t+\phi_2)+A^\times\sin\left(\omega_2t+\phi_2\right)\right]\\
  =&\mu_s\Big[A^+\cos\left(\omega_\text{f}t+\phi_\text{f}\right)\cos\left(\omega_\text{b}t+\phi_\text{b}\right)\\
   &+A^\times\cos\left(\omega_\text{f}t+\phi_\text{f}-\frac{\pi}{2}\right)\cos\left(\omega_\text{b}t+\phi_\text{b}\right)\Big]+\\
  &\mu_d\left[A^+\cos\left(\omega_\text{f}t+\phi_\text{f}+\frac{\pi}{2}\right)\cos\left(\omega_\text{b}t+\phi_\text{b}-\frac{\pi}{2}\right)\right.\\
  &\left.+A^\times\cos\left(\omega_\text{f}t+\phi_\text{f}\right)\cos\left(\omega_\text{b}t+\phi_\text{b}-\frac{\pi}{2}\right)\right],
 \end{split}
\end{equation}
where $A^{+/\times}$ stand for the amplitudes of the rays, $\mu_s=\mu_++\mu_-$ and $\mu_d=\mu_+-\mu_-$.
If these rays were generated by a binary system with masses $m_1$ and $m_2$ circling around each other,
their amplitudes are roughly,
\begin{gather}
  A^+=\frac{4\mathcal M}{d_\text{L}}(\pi\mathcal Mf)^{2/3}F^+\frac{1+\cos^2\iota}{2},\label{eq-as-p}\\
  A^\times=\frac{4\mathcal M}{d_\text{L}}(\pi\mathcal Mf)^{2/3}F^\times\cos\iota,\label{eq-as-c}
\end{gather}
at the leading order, where $\mathcal M=(1+z_\text{S})(m_1m_2)^{3/5}/(m_1+m_2)^{1/5}$ is the redshifted chirp mass,  $d_\text{L}$ is the luminosity distance, $\iota$ is the inclination angle and $F^{+/\times}$ are the antenna pattern functions \cite{Isi:2017fbj,Schilling:1997id,*Giampieri:1997kv,*Liang:2019pry}.
In addition, $\omega_1$ and $\omega_2$ in Eq.~\eqref{eq-s-12} are the angular frequencies of the GWs, $\phi_1$ and $\phi_2$ are their initial phases, and
\begin{gather}
\omega_\text{f}=\frac{\omega_1+\omega_2}{2},\quad\phi_\text{f}=\frac{\phi_1+\phi_2}{2},\\
\omega_\text{b}=\frac{\omega_1-\omega_2}{2},\quad\phi_\text{b}=\frac{\phi_1-\phi_2}{2}.
\end{gather}
In the early inspiral phase, $2\omega_\text{b}$ can be much smaller than $2\omega_\text{f}$, because of the small time delay $\Delta t$, so the beat pattern forms in the time domain with the beat frequency $\omega_\text{b}$.


For the binary star system during the inspiral phase, the GW angular frequency evolves according to \cite{Maggiore:1900zz}
\begin{equation}\label{eq-t-eom}
  \frac{\ud\omega}{\ud t}=\frac{192}{5}\mathcal M^{5/3}\left( \frac{\omega}{2} \right)^{11/3},
\end{equation}
at the leading order.
When $\omega_\text{b}\ll\omega_\text{f}$, Eq.~\eqref{eq-t-eom} can be used to calculate
\begin{equation}\label{eq-ombt}
  \omega_\text{b}\approx\frac{96}{5}\left(\frac{\omega_\text{f}}{2}\right)^{11/3}\mathcal M^{5/3}\Delta t,
\end{equation}
and $\omega_\text{f}$ is roughly the GW angular frequency.
So, as the time advances, $\omega_\text{f}$ increases, and $\omega_\text{b}/\omega_\text{f}\propto\omega_\text{f}^{8/3}$ becomes larger.
Eventually, the beat pattern disappears.
After the compact-object merger is observed along ray 1, its strain drops dramatically, and the total strain is basically $h\approx h_2$, which has the same behavior as usual waveform with the amplitude magnified by a factor of ${\mu_-}$.

\begin{figure}
  \centering
  \includegraphics[width=.45\textwidth]{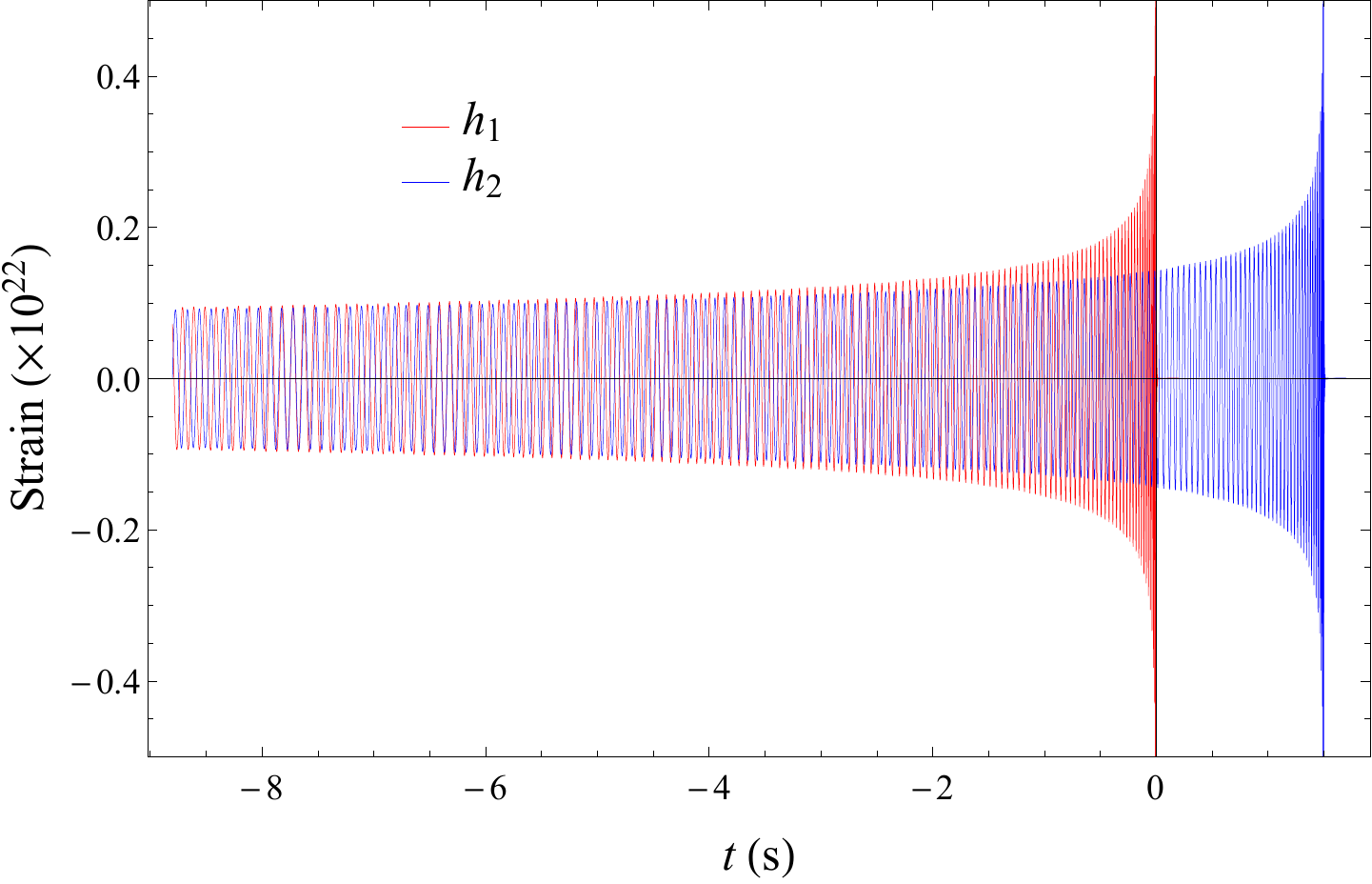}
  \includegraphics[width=.45\textwidth]{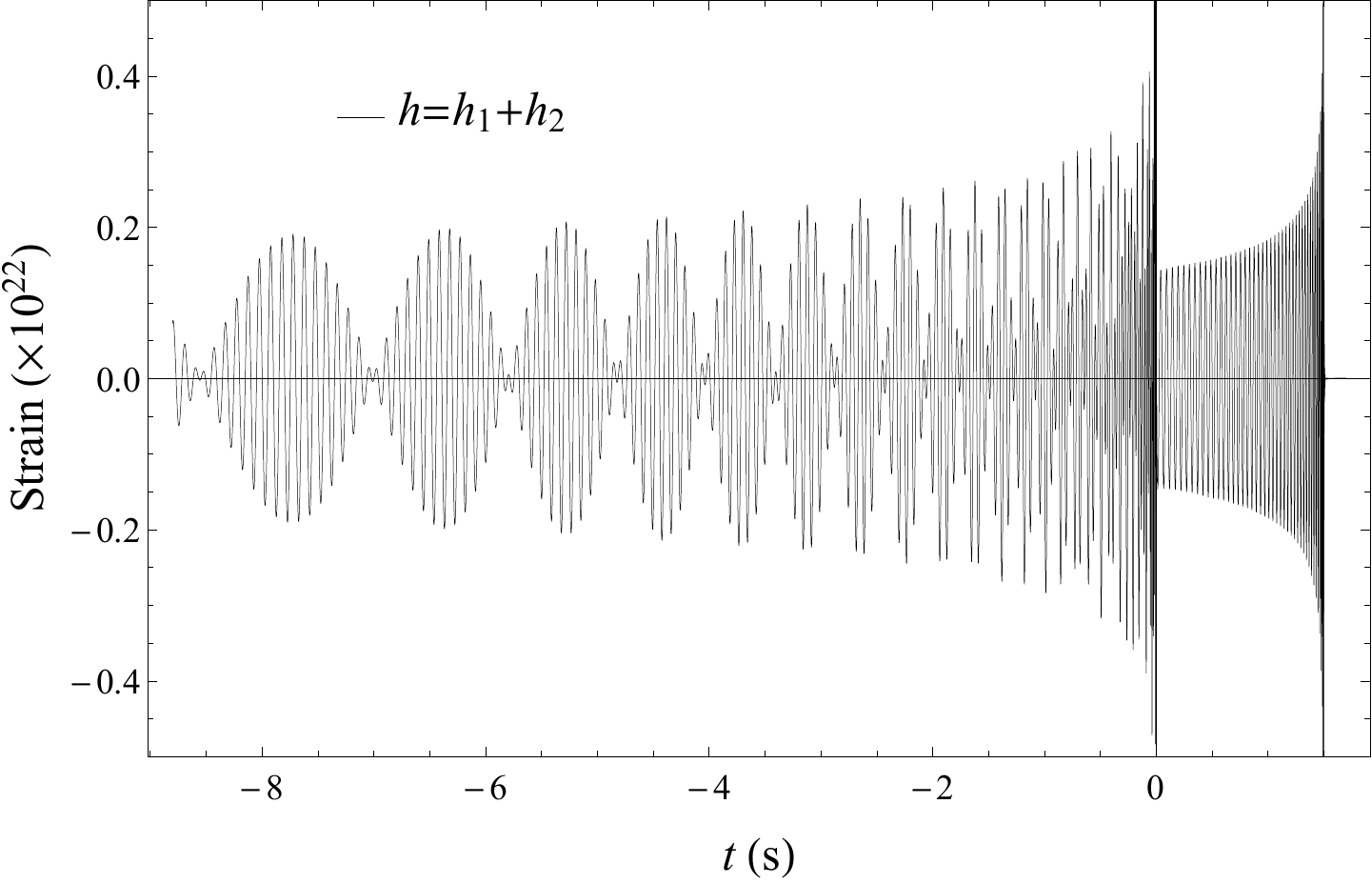}
  \caption{The schematic diagrams showing the strains and  the beat pattern in the time domain.
  In the two panels, the horizontal axes are time $t$ in seconds, and the vertical axes are strain $10^{22}h$.
}\label{fig-int}
\end{figure}

With PyCBC \cite{Canton:2014ena,*Usman:2015kfa,*alex_nitz_2019_2643618s}, one can easily simulate the interference between the two GW rays 1 and 2.
For example, let $m_1=30M_\odot$ and $m_2=20M_\odot$, and the binary system is assumed to be at $z_\text{S}=2$.
Suppose the lens is a point mass with $M=10^{6}M_\odot$ and is at $z_\text{L}=1$.
Choose $\beta=2.5\times10^{-5}$ arc sec; then the magnification factors are ${\mu_+}\approx2.27$ and ${\mu_-}\approx2.25$.
With these choices of parameters, the time delay is $\Delta t\approx1.49$ sec, which is small enough.
Although $\beta$ is very small, the wave optics can still be ignored because the time delay $\Delta t$ is still much longer than the period of the GW in the detector bands \cite{Nakamura1999wo}.
For the purpose of demonstration, we assume that the two GR rays are parallel to each other as the deflection angles are $\alpha_1\approx\alpha_2\approx3.6\times10^{-3}$ arc sec.
The inclination angle is chosen to be $\iota=0$.
Let the GWs travel in the direction perpendicular to the detector arms; then the strains $h_1,\,h_2$, and $h$ are shown in Fig.~\ref{fig-int}.
The upper panel shows the strains $h_1$ and $h_2$, separately.
The lower panel shows the total strain $h=h_1+h_2$.
At the earlier time, e.g., $t\lesssim-2$ sec, there exists a fairly good beat pattern whose frequency $\omega_\text{b}$ increases with time.
After $-2$ sec, the beat pattern gets disturbed due to the growth in $\omega_\text{b}/\omega_\text{f}$, and finally, it disappears.
Although Fig.~\ref{fig-int} is for the binary black hole merger, it is easy to understand that similar waveforms also apply to the binary neutron star and neutron star-black hole mergers.

The lower panel in Fig.~\ref{fig-int} depicts the characteristic feature of the beat pattern.
As long as one observes such a kind  of events, one knows that the GWs were probably lensed \cite{Nakamura1999wo}.
Several quantities can be determined utilizing the lensed waveform.
The beat frequency $\omega_\text{b}$ and the average frequency $\omega_\text{f}$ can be read off from the waveform in the early inspiral stage.
The chirp mass $\mathcal M$ is determined from the strain after the merger is observed through the first ray, and thus Eq.~\eqref{eq-ombt} gives $\Delta t$, which could also be read off from Fig.~\ref{fig-int} by measuring the length from the first merge to the second.
Therefore, the time delay can be determined without the aid of the electromagnetic counterpart in principle, with a \emph{higher} precision on the order of $\sim0.1$ sec.

\subsection{Signal-to-noise ratio}
\label{sec-snr-g}

\begin{figure}[h]
  \centering
  \includegraphics[width=0.45\textwidth]{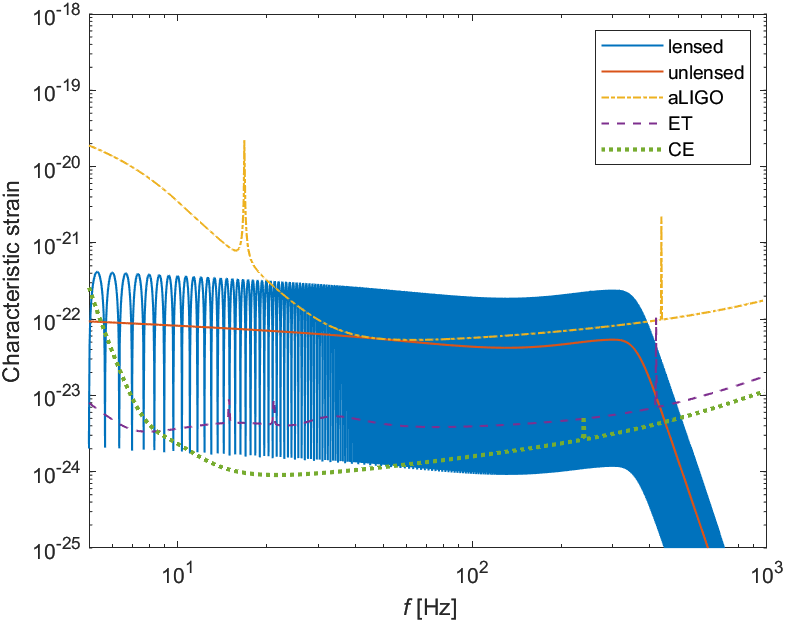}
  \caption{The characteristic strains for the lensed GW (solid blue curve) and the unlensed GW (solid brown curve) together with the sensitivity curves of aLIGO (dot-dashed yellow curve), ET (dashed purple curve) and CE (dotted greeen curve).
  The sensitivity curves are taken from PyCBC \cite{Canton:2014ena,*Usman:2015kfa,*alex_nitz_2019_2643618s}.
  }
  \label{fig-gr}
\end{figure}

The frequency domain waveform for the beat is also easy to obtain and is used to calculate the SNR \cite{Yunes:2013dva}.
Let $\tilde h_1(f)$ be the frequency domain waveform for $h_1(t)$, i.e.,
\begin{equation}
  \label{eq-h1f}
  \begin{split}
  \tilde{h}_1(f)&=\mu_+\int_{-\infty}^\infty h_u(t)e^{i2\pi ft}\ud f\\
    =&\mu_+\tilde h_u(f),
  \end{split}
\end{equation}
where $h_u(t)$ is for the unlensed waveform and $\tilde h_u(f)$ is its Fourier transform; then the frequency domain waveform $\tilde h_2(f)$ for $h_2(t)=\frac{\mu_-}{\mu_+}h_1(t-\Delta t)$ is 
\begin{equation}
  \label{eq-h2f}
  \tilde h_2(f)=e^{i2\pi f\Delta t}\mu_-\tilde h_u(f).
\end{equation}
So the amplitude of the total waveform is 
\begin{equation}
  \label{eq-hf}
  |\tilde h(f)|=\sqrt{\mu_+^2+\mu_-^2+2\mu_+\mu_-\cos(2\pi f\Delta t)}|\tilde h_u(f)|.
\end{equation}
Figure~\ref{fig-gr} shows the characteristic strains of the lensed and the unlensed signals, together with the sensitivity curves of aLIGO, Einstein Telescope (ET) and Cosmic Explorer (CE) \cite{Punturo:2010zza,Evans:2016mbw}.
It shows that the beat strain is highly oscillating in the frequency domain, due to the small $\Delta t$. 
In fact, the ``period'' of this oscillation is $1/\Delta t\approx0.67$ Hz.
\begin{table}
  \bgroup
  \setlength\tabcolsep{0.03\textwidth}
  \begin{tabular}
    {c|ccc}
    \hline\hline
    & aLIGO & ET & CE \\ 
    \hline 
    Lensed & 4.09 & 78.0 & 294 \\
    \hline
    Unlensed & 1.28 & 24.3 & 91.8\\ 
    \hline\hline
  \end{tabular}
  \egroup
  \caption{SNRs for the GW considered in Fig.~\ref{fig-int}.
  The detectors are chosen to be aLIGO, ET and CE.
  The unlensed SNRs correspond to GW signals shaped like the red and blue curves but lack the magnification factors $\mu_\pm$.
  }
  \label{tab-snrs-g}
\end{table}

The SNR for this signal $\tilde h(f)$ is given by \cite{Mirshekari:2011yq}
\begin{equation}
  \label{eq-snr-def}
  \rho=2\left(\int_0^\infty\frac{|\tilde h(f)|^2}{S_n(f)}\right)^{1/2},
\end{equation}
with $S_n(f)$ being the noise power spectrum for a detector, which can be generated using PyCBC. 
The SNRs for the signals presented in Fig.~\ref{fig-int} can thus be calculated, assuming the detector is aLIGO, ET, or CE, and tabulated in Table~\ref{tab-snrs-g}. 
In this table, ``unlensed'' refers to the SNR calculated for the red or the blue signal in the upper panel of Fig.~\ref{fig-int}, while ``lensed'' is for the  signal in the lower panel.
It is clear that the interference causes increases in SNRs due to  i) the amplification of the individual GW signal and ii) the simultaneous detection of them.
In calculating the SNRs, we assume that the exactly correct template is used.
Therefore, with the suitable templates, the lensed signals can be detected by ET or CE.

\subsection{Application}
\label{sec-b-app}

Suppose such a kind of signal can be detected with a high SNR; then one can make use of it to extract some useful information about the lens and even some cosmological parameters.
By Eq.~\eqref{eq-s-12}, it is possible to infer $\mu_s/\mu_d$ by matched filtering once the template for the lensed GWs is used.
If $\mu_d$ is too small to be determined, one can still obtain $\mu_s/{\mu_-}$ by measuring the amplitude of $h_2$ using the data after the merger is detected through the first ray.
So the relative magnification ${\mu_+/\mu_-}$ is known, 
\begin{equation}
  \label{eq-rmus}
  \frac{\mu_+}{\mu_-}=\frac{1+\mu_s/\mu_d}{1-\mu_s/\mu_d}=\frac{\mu_s}{\mu_-}-1,
\end{equation}
which can be used to find the ratio $r_\pm=\theta_+/\theta_-=\mu_+/\mu_-$ with Eq.~\eqref{eq-mf}.
Substituting $r_\pm$ back into Eq.~\eqref{eq-mf}  gives the actual magnifications, 
\begin{equation}
  {\mu_+}=\frac{1}{\sqrt{\left|1-r_\pm^{-2}\right|}},\quad {\mu_-}=\frac{1}{\sqrt{\left|r_\pm^2-1\right|}}.
\end{equation}
With these, one can unambiguously calculate the luminosity distance of the GW source.
Further, using $r_\pm$, Eqs.~\eqref{eq-dan} and \eqref{eq-mf}, one finds  $\beta/\theta_\text{E}$, which can be substituted into Eq.~\eqref{eq-tdv}, with the help of Eq.~\eqref{eq-dan}, to find the redshifted lens mass,
\begin{equation}
  \label{eq-slm}
  M(1+z_\text{L})=\frac{\Delta t}{4}\left[ \frac{1}{2}\left|r_\pm-\frac{1}{r_\pm}\right|+\ln(-r_\pm) \right]^{-1}.
\end{equation}

The above application relies on the observation of several beats, so the observation time should be a few seconds, which can be achieved for the third-generation detectors such as ET and CE.
One can estimate how long an inspiral signal lasts given an initial frequency by integrating Eq.~\eqref{eq-t-eom} to get the GW frequency,
\begin{equation}
  \label{eq-ft}
  f(t)=\left[f_c^{-8/3}-\frac{256\pi}{5}(\pi\mathcal M)^{5/3}(t-t_c)\right]^{-3/8},
\end{equation}
and then solving for $t-t_c$.
Here, $t_c$ is the fiducial coalescence time, and $f_c$ is the corresponding coalescence frequency.
One may choose $f_c=f_\text{isco}$ with \cite{Bonvin:2016qxr}
\begin{equation}
  \label{eq-fisco}
  f_\text{isco}=8.80(1+1.25\eta+1.08\eta^2)\left[ \frac{M_\odot}{(1+z_\text{S})(m_1+m_2)} \right]\text{kHz},
\end{equation}
where $\eta=m_1m_2/(m_1+m_2)^2$ is the symmetric mass ratio.
This is the GW frequency when the innermost stable circular orbit  of the binary system is reached.
Take the unlensed GW signal in the previous subsections, for example. 
From Fig.~\ref{fig-gr}, the unlensed signal is well above the sensitivity curves of ET and CE from 5 Hz; then using the above equation, one finds out that the time left to reach $f_\text{isco}$ is about 8.6 sec, which is long enough to contain several beat periods. 

This application also calls for high enough lensing event rates.
Several works discuss the lensing rates for LIGO \cite{Ng:2017yiu} and ET \cite{Ding:2015uha,Li:2018prc,Yang:2019jhw}.
At the design sensitivity, aLIGO could only detect about five lensed GW events per year, while the event rate increases to about 80 to 100 per year for ET.
In their calculations, some more realistic lens models such as SIS were considered, and the lensed GW events were required to have high enough SNRs ($\ge8$).
The event rate for the beat pattern would be even smaller, because of the extreme restriction on $\Delta t$, or $\beta$. 
Although the rate will be calculated in future work, here we roughly estimate it. 
Assuming the SIS model (in order to be more realistic), $\beta$ should be around $10^{-7}$ arc sec if the source and the lens are still at $z_\text{S}=2$ and $z_\text{L}=1$, respectively, and $\sigma=250$ km/s \cite{Fan:2016swi}.
The Einstein angle is $\theta_\text{E}=0.65$ arc sec, so $\beta/\theta_\text{E}\approx10^{-6}\sim10^{-7}$.
Then by Ref.~\cite{Piorkowska:2013eww}, the lensing cross section and thus the optical depth would be about 12 to 14 orders of magnitude smaller than those considered in Refs.~\cite{Ding:2015uha,Ng:2017yiu,Li:2018prc,Yang:2019jhw}, so the lensing rate for the beat is much less than the predicted values quoted above.

Finally, the possibilities of obtaining the actual magnification factors and the redshifted lens mass are due to the choice of the lens model, which gives simple relations \eqref{eq-mf} and \eqref{eq-tdv}. 
In reality, the lens density profile is complicated and the relations like \eqref{eq-mf} and \eqref{eq-tdv} might not be simple enough to use for inferring useful information. 

For the space-borne interferometers, the restriction on $\beta$ is very relaxed, and the lensing rate is expected to be higher.
So in the next section, we will discuss the beat detectable by these detectors.

\section{Interference observed by the space-borne detectors}
\label{sec-ios}

From the above discussion, one finds out that in order to detect the beat pattern on the ground there is a tremendously tight bound on $\beta\sim 10^{-5}-10^{-7}$ arc sec.
This leads to the extremely low probabilities of observing beat pattern even with ET or CE.
The tight bound is due to the small observation periods available to these detectors. 
In contrast, the target GWs of the space-borne interferometers could last for much longer amounts of time. 
The formation of the beat pattern by lensing these GWs places milder constraint on $\beta$.
So, it is useful to study the beat pattern observed by detectors such as LISA and the application.

The following discussion will parallel to that in Sec.~\ref{sec-iog}.
To avoid the repetition, we will emphasize some of the differences, which are mainly the less restricted misalignment angle $\beta$, the longer and more beat periods. 

\subsection{Beat and its SNR} 
\label{sec-beat-s}

Before demonstrating a concrete example of a beat pattern, we should notice that LISA, for instance, is sensitive in frequency range $10^{-4}-10^{-1}$ Hz. 
So, one should expect that the beat period can be very long, as the beat angular frequency $\omega_\text{b}\ll\omega_\text{f}$ with $\omega_\text{f}$ roughly the GW angular frequency. 
If the beat period is too long, e.g., on the order of a few months, then the orbital motion of the LISA satellites would have some impact on the interference pattern detected by LISA. 
This is due to the fact that the antenna pattern functions $F^{+/\times}$ [refer to Eqs.~\eqref{eq-as-p} and \eqref{eq-as-c}] depend on the relative orientation of the GW to the constellation plane of LISA. 
So, for a fixed GW source, LISA would register different strains at different positions in its orbit \cite{Liang:2019pry}, as nicely demonstrated in  Ref.~\cite{gong_yungui_2019_2574620}.
This effect would make the analysis of the beat pattern more complicated, since $F^{+/\times}$ are effectively also functions of time $t$. 
However, $\omega_\text{b}$ grows over time by Eq.~\eqref{eq-ombt}.
Thus, there is the possibility that $\omega_\text{b}$ becomes large enough that the beat period is on the order of a few hours to a few minutes.
When this happens, the orbital motion of LISA might be ignored. 
Although the varying $F^{+/\times}$ during the long beat periods do not fundamentally change the analysis of the beat, in the current work, we focus on the simpler case in which the beat periods are small enough to ignore the orbital motion of LISA.

To make sure that there exists some time window when the beat periods are small enough to ignore the orbital motion of LISA, the time delay $\Delta t$ should be big enough.
But at the same time, $\Delta t$ should also be bounded from above; otherwise $\omega_\text{b}$ might quickly increase to such a good portion of $\omega_\text{f}$ in a relatively short time that there are not enough beats for us to use. 
In the following, we will exhibit a binary system and a SIS lens to satisfy these conditions.

\begin{figure}[h]
  \centering
  \includegraphics[width=.45\textwidth]{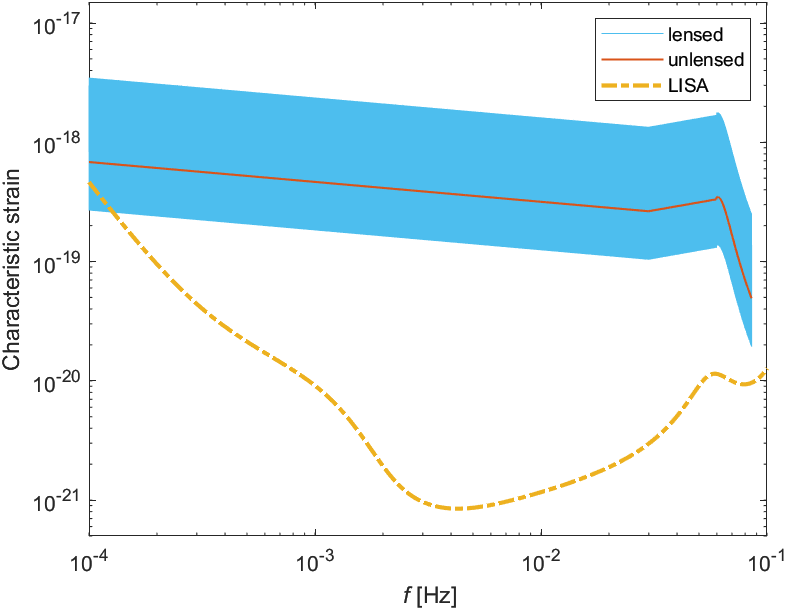}
  \caption{The characteristic strains for the lensed GW (solid cyan curve), the unlensed GW (solid brown curve), and the sensitivity curve of LISA (dot-dashed orange curve).
  This plot is generated using the LISA sensitivity calculator \cite{neil_cornish_2019_2636514}.}
  \label{fig-sb}
\end{figure}

The suitable sources of GWs detectable by the space-borne interferometers usually have much heavier masses than the above example.
For example, one may consider a binary star system of masses $5\times10^4M_\odot$ and $4\times10^4M_\odot$ at the redshift $z_\text{S}=2$.
Let the lens be an early-type galaxy with the velocity dispersion $\sigma=250\text{ km/s}$ \cite{Fan:2016swi}.
Then set $\beta=0.1$ arc sec, resulting in $\theta_\text{E}=0.65$ arc sec,  $\mu_+\approx2.75$, and $\mu_-\approx2.35$.
The time delay is $\Delta t=1.05$ months, which is an appropriate value.
Here, we display the frequency domain waveforms for the beat and the unlensed GW in Fig.~\ref{fig-sb}, together with the sensitivity curve of LISA, using the LISA sensitivity calculator \cite{Cornish:2018dyw,neil_cornish_2019_2636514}.
In this plot, the unlensed GW is represented by the solid brown curve, which is well above the sensitivity curve of LISA, the dot-dashed orange curve. 
One can easily calculate its SNR, which is approximately 517.
The cyan curve is for the interfered signal. 
According to Eq.~\eqref{eq-hf}, the strain of the beat $\tilde h(f)$ is oscillating in the frequency domain with the period $1/\Delta t\approx3.66\times10^{-7}$ Hz.
Such a small period makes the cyan curve curl up to form a band. 
Its SNR is about 1870.
 
Suppose this interfered signal can be detected from $10^{-4}$ Hz; then, it can be estimated that $f_\text{isco}$ is reached after nearly 1.36 years. 
Integrating Eq.~\eqref{eq-ft} over time gives the phase of the GW,
\begin{equation}
  \label{eq-gwph}
  \begin{split}
  \phi(t)=&\phi_c+\frac{1}{16}\left\{\frac{1}{(\pi\mathcal Mf_c)^{5/3}}\right.\\
  &\left.-\left[ \frac{1}{(\pi\mathcal Mf_c)^{8/3}}-\frac{256}{5}\frac{t-t_c}{\mathcal M} \right]^{5/8}\right\},
  \end{split}
\end{equation}
where $\phi_c$ is a fiducial coalescence phase.
This facilitates the determination of the beat period more accurately.
In fact, the phase for the beat is $\phi_\text{b}(t)=\phi(t)-\phi(t-\Delta t)$.
Now, set $\cos\phi_\text{b}(t)=0$; then, one can solve this equation to determine the time $t_n$ when $\phi_\text{b}(t_n)=(n+1/2)\pi$ with $n$ some integer, and $t_{n+1}-t_n$ is roughly half of the beat period.
It turns out that the periods for the first few beats are from about two weeks down to one or two days.
So none of these beats can be used to easily extract useful information. 
There are actually some beats with small enough periods, on the order of a few hours.
These beats start from the time when there are just 10.6 months before the GW frequency reaches $f_\text{isco}$.
The last one of them occurs about 1.49 months before $f_\text{isco}$.
In total, there are 407 such a kind of beats with the frequency $\omega_\text{f}/2\pi$ ranging from $f_1=1.66\times10^{-4}$ to $f_2=3.48\times10^{-4}$ Hz, so it is possible to use them to extract some useful information as done in Sec.~\ref{sec-b-app}.

\subsection{Application}
\label{sec-be-s-app}

With the relations presented in the previous sections, one concludes that it is still possible to obtain the magnification factors.
First, by Eq.~\eqref{eq-ang-sis}, $\frac{\theta_+}{\theta_\text{E}}-\frac{\theta_-}{\theta_\text{E}}=2$. 
Once the ratio $\alpha=\mu_+/\mu_-$ is inferred using the matched filtering as given by Eq.~\eqref{eq-rmus},  these give
\begin{equation}
  \label{eq-ys}
  \frac{\theta_+}{\theta_\text{E}}=\frac{2\alpha^2}{\alpha^2+1},\quad \frac{\theta_-}{\theta_\text{E}}=\frac{-2}{\alpha^2+1}.
\end{equation}
These can be substituted back into Eq.~\eqref{eq-mu-sis} to calculate $\mu_\pm$, 
\begin{equation}
  \label{eq-ma}
  \mu_+=\sqrt{\frac{2\alpha^2}{\alpha^2-1}},\quad
   \mu_-=\sqrt{\frac{2}{\alpha^2-1}}.
\end{equation}
Therefore, the actual luminosity distance of the source is also determined.

As discussed above, the time delay $\Delta t$ can be measured by two different methods.
One is to simply find the times of the two mergers and take their difference, and the other is to measure the beat frequency $\omega_\text{b}$, which is substituted into Eq.~\eqref{eq-ombt} to solve for $\Delta t$.
The second method is better in the sense that some GW signals merge at the higher frequencies beyond LISA's sensitivity band.
If the lensing galaxy can be identified and its redshift $z_\text{L}$ can be determined (e.g., via follow-up imaging or spectroscopic observations as suggested by Ref.~\cite{Sereno:2011ty}), Eq.~\eqref{eq-tdv-sis} gives the time-delay distance $D_{\Delta t}=D_\text{L}D_\text{LS}/D_\text{S}$, 
\begin{equation}
  \label{eq-ddt}
  D_{\Delta t}=\frac{\Delta t}{32\pi^2\sigma_v^2(1+z_\text{L})}\frac{\alpha^2+1}{\alpha^2-1}.
\end{equation}
In the Friedmann-Robertson-Walker spacetime, the angular diameter distance is given by \cite{Weinberg:2008zzc},
\begin{equation}
  \label{eq-daw}
  \begin{split}
  D(z)&=\frac{1}{\sqrt{\Omega_k}H_0(1+z)}\times\\ 
  &\sinh\left[ \sqrt{\Omega_k}\int_{\frac{1}{1+z}}^1\frac{\ud x}{\sqrt{\Omega_r+\Omega_mx+\Omega_kx^2+\Omega_\Lambda x^4}} \right],
  \end{split}
\end{equation}
where $H_0$ is the present-day Hubble constant and $\Omega_l$ are the density parameters for the radiation ($l=r$), the matter ($l=m$), the spatial curvature ($l=k$), and the dark energy ($l=\Lambda$).
Therefore, $D_{\Delta t}$ could be used to constrain the cosmological parameters, especially the Hubble constant $H_0$ \cite{Liao:2017ioi,Liu:2019dds}.
For this purpose, one has to use the electromagnetic follow-up observations to determine the host galaxy of the GW source and thus $z_\text{S}$ \cite{Holz:2005df}.
The advantage of using the beat pattern is that one need not find all the ``images'' in order to determine the Fermat potential.
So, this method would be more accurate.

The conservative estimation of the lensing rate for LISA has been done in Ref.~\cite{Sereno:2010dr}.
During a five-year mission, there can be 4 multiple events detectable with SNRs $\ge8$.
From the previous subsection, one finds out that in order to detect the beat pattern using LISA $\beta$ can be on the order of 0.1 arc sec, which is not very restricted.
In this case, $\beta/\theta_\text{E}\approx0.15$, which, compared with $y_\text{max}$ in Fig.~1 in Ref.~\cite{Sereno:2010dr}, does not have the difference on a few orders of magnitude.
So, it is possible to observe the beat with LISA. 
More exact calculation of the event rate for LISA or DECIGO will be done in future work. 

In the end, one should also realize that Eqs.~\eqref{eq-ma} and \eqref{eq-ddt} are obtained based on \eqref{eq-mu-sis} and \eqref{eq-tdv-sis}, because the SIS model is used in this section. 
Although the SIS model is more realistic compared with the point mass model in Sec.~\ref{sec-beat-g}, it is still very ideal and does not fully reflect the complexities of the real lenses. 
Therefore, as stated in the final part of Sec.~\ref{sec-b-app}, a real lens model may not allow one to extract this useful information.

\section{Discussion and Conclusion} 
\label{sec-con}

As discussed above, when the time delay between the lensed GWs is small enough, it is possible to obtain the actual luminosity distance by making good use of the beat pattern.
The time delay $\Delta t$ can be easily measured at a very high precision without the need for the electromagnetic counterpart.
The gravitational lensing caused by a massive black hole, modeled as a point mass, also enables the measurement of the redshifted mass of the lens, which does not rely on any statistics method, but the event rate is negligible.
Since space-borne interferometers have much longer observation periods, there would be more GW events with the beat pattern detected, and the method presented above could put constraints on the cosmological parameters, if the lens is described by a SIS.
All of the above applications of the beat pattern rely on the fact that the lens models considered are simple enough.

In Secs.~\ref{sec-snr-g} and \ref{sec-beat-s}, the SNRs are calculated in order to show that there can be lensed GWs that are louder than the detector noise. 
However, in order to know whether it is possible to extract useful information from the lensed GW signal, such as the beat frequency $\omega_\text{b}$, one has to conduct more detailed analysis to decide the threshold SNR above which $\omega_\text{b}$ can be measured accurately enough. 
In addition, in the actual application of the beat pattern, one has to take into account several factors that limit the use of this interference for cosmology. 
These factors include, but are not limited to, the finite lensing event rates, the finite SNRs, the intrinsic scatter in the lens profile and the cosmic shear. 
These factors make it difficult to measure some cosmological parameters accurately. 
The effects of some of them might be partially mitigated, as discussed in Sec.~\ref{sec-int} (about microlensing and weak lensing).
The total number of lensed GW events as predicted by Ref.~\cite{Sereno:2010dr} is  small, if LISA is used. 
But the lensing rate might be much higher, if DECIGO  is used.
This is because DECIGO is mostly sensitive to GWs in the decihertz range, which allows higher beat frequencies $\omega_\text{b}$, and the orbital motion can be largely ignored. 
So, in principle, one might observe more beats and reduce the systematic errors. 
The higher working frequencies also imply that DECIGO could detect GWs emitted by smaller binary systems more easily.
The merger rate of the smaller binary systems is much higher than that of the massive black hole binaries  \cite{Sesana:2008ur,Klein:2015hvg,LIGOScientific:2018mvr,Gerosa:2019dbe,Boco_2019}, so DECIGO could potentially detect a lot more lensed GWs, which might allow us to use the statistical methods  to constrain some measurable quantities discussed previously. 
Since these GW  events would last for a long time, also on the order of years, it is expected that most of the lensed GWs would interfere with each other and form  the beat patterns. 
Finally, with the long observation periods and the higher sensitivity, the SNRs might be large enough.
Nevertheless, the limiting factors have to be properly taken into account in the actual analysis.

\begin{acknowledgements}
  We would like to thank Hai Yu for the constructive discussions.
  This work was supported by the National Natural Science Foundation of China under Grants No.~11633001, No.~11673008, No.~11922303, and No.~11920101003 and the Strategic Priority Research Program of the Chinese Academy of Sciences, Grant No. XDB23000000.
  Xilong Fan is supported by Hubei province Natural Science Fund for the Distinguished Young Scholars.
\end{acknowledgements}

\end{CJK*}
\bibliographystyle{apsrev4-1}
\bibliography{beatprd.bbl}

\begin{thebibliography}{86}%
\makeatletter
\providecommand \@ifxundefined [1]{%
 \@ifx{#1\undefined}
}%
\providecommand \@ifnum [1]{%
 \ifnum #1\expandafter \@firstoftwo
 \else \expandafter \@secondoftwo
 \fi
}%
\providecommand \@ifx [1]{%
 \ifx #1\expandafter \@firstoftwo
 \else \expandafter \@secondoftwo
 \fi
}%
\providecommand \natexlab [1]{#1}%
\providecommand \enquote  [1]{``#1''}%
\providecommand \bibnamefont  [1]{#1}%
\providecommand \bibfnamefont [1]{#1}%
\providecommand \citenamefont [1]{#1}%
\providecommand \href@noop [0]{\@secondoftwo}%
\providecommand \href [0]{\begingroup \@sanitize@url \@href}%
\providecommand \@href[1]{\@@startlink{#1}\@@href}%
\providecommand \@@href[1]{\endgroup#1\@@endlink}%
\providecommand \@sanitize@url [0]{\catcode `\\12\catcode `\$12\catcode
  `\&12\catcode `\#12\catcode `\^12\catcode `\_12\catcode `\%12\relax}%
\providecommand \@@startlink[1]{}%
\providecommand \@@endlink[0]{}%
\providecommand \url  [0]{\begingroup\@sanitize@url \@url }%
\providecommand \@url [1]{\endgroup\@href {#1}{\urlprefix }}%
\providecommand \urlprefix  [0]{URL }%
\providecommand \Eprint [0]{\href }%
\providecommand \doibase [0]{http://dx.doi.org/}%
\providecommand \selectlanguage [0]{\@gobble}%
\providecommand \bibinfo  [0]{\@secondoftwo}%
\providecommand \bibfield  [0]{\@secondoftwo}%
\providecommand \translation [1]{[#1]}%
\providecommand \BibitemOpen [0]{}%
\providecommand \bibitemStop [0]{}%
\providecommand \bibitemNoStop [0]{.\EOS\space}%
\providecommand \EOS [0]{\spacefactor3000\relax}%
\providecommand \BibitemShut  [1]{\csname bibitem#1\endcsname}%
\let\auto@bib@innerbib\@empty
\bibitem [{\citenamefont {Abbott}\ \emph
  {et~al.}(2016{\natexlab{a}})\citenamefont {Abbott} \emph
  {et~al.}}]{Abbott:2016blz}%
  \BibitemOpen
  \bibfield  {author} {\bibinfo {author} {\bibfnamefont {B.~P.}\ \bibnamefont
  {Abbott}} \emph {et~al.} (\bibinfo {collaboration} {Virgo, LIGO
  Scientific}),\ }\href {\doibase 10.1103/PhysRevLett.116.061102} {\bibfield
  {journal} {\bibinfo  {journal} {Phys. Rev. Lett.}\ }\textbf {\bibinfo
  {volume} {116}},\ \bibinfo {pages} {061102} (\bibinfo {year}
  {2016}{\natexlab{a}})},\ \Eprint {http://arxiv.org/abs/1602.03837}
  {arXiv:1602.03837 [gr-qc]} \BibitemShut {NoStop}%
\bibitem [{\citenamefont {Abbott}\ \emph
  {et~al.}(2016{\natexlab{b}})\citenamefont {Abbott} \emph
  {et~al.}}]{Abbott:2016nmj}%
  \BibitemOpen
  \bibfield  {author} {\bibinfo {author} {\bibfnamefont {B.~P.}\ \bibnamefont
  {Abbott}} \emph {et~al.} (\bibinfo {collaboration} {Virgo, LIGO
  Scientific}),\ }\href {\doibase 10.1103/PhysRevLett.116.241103} {\bibfield
  {journal} {\bibinfo  {journal} {Phys. Rev. Lett.}\ }\textbf {\bibinfo
  {volume} {116}},\ \bibinfo {pages} {241103} (\bibinfo {year}
  {2016}{\natexlab{b}})},\ \Eprint {http://arxiv.org/abs/1606.04855}
  {arXiv:1606.04855 [gr-qc]} \BibitemShut {NoStop}%
\bibitem [{\citenamefont {Abbott}\ \emph
  {et~al.}(2017{\natexlab{a}})\citenamefont {Abbott} \emph
  {et~al.}}]{Abbott:2017vtc}%
  \BibitemOpen
  \bibfield  {author} {\bibinfo {author} {\bibfnamefont {B.~P.}\ \bibnamefont
  {Abbott}} \emph {et~al.} (\bibinfo {collaboration} {Virgo, LIGO
  Scientific}),\ }\href {\doibase 10.1103/PhysRevLett.118.221101} {\bibfield
  {journal} {\bibinfo  {journal} {Phys. Rev. Lett.}\ }\textbf {\bibinfo
  {volume} {118}},\ \bibinfo {pages} {221101} (\bibinfo {year}
  {2017}{\natexlab{a}})},\ \Eprint {http://arxiv.org/abs/1706.01812}
  {arXiv:1706.01812 [gr-qc]} \BibitemShut {NoStop}%
\bibitem [{\citenamefont {Abbott}\ \emph
  {et~al.}(2017{\natexlab{b}})\citenamefont {Abbott} \emph
  {et~al.}}]{Abbott:2017oio}%
  \BibitemOpen
  \bibfield  {author} {\bibinfo {author} {\bibfnamefont {B.~P.}\ \bibnamefont
  {Abbott}} \emph {et~al.} (\bibinfo {collaboration} {Virgo, LIGO
  Scientific}),\ }\href {\doibase 10.1103/PhysRevLett.119.141101} {\bibfield
  {journal} {\bibinfo  {journal} {Phys. Rev. Lett.}\ }\textbf {\bibinfo
  {volume} {119}},\ \bibinfo {pages} {141101} (\bibinfo {year}
  {2017}{\natexlab{b}})},\ \Eprint {http://arxiv.org/abs/1709.09660}
  {arXiv:1709.09660 [gr-qc]} \BibitemShut {NoStop}%
\bibitem [{\citenamefont {Abbott}\ \emph
  {et~al.}(2017{\natexlab{c}})\citenamefont {Abbott} \emph
  {et~al.}}]{TheLIGOScientific:2017qsa}%
  \BibitemOpen
  \bibfield  {author} {\bibinfo {author} {\bibfnamefont {B.~P.}\ \bibnamefont
  {Abbott}} \emph {et~al.} (\bibinfo {collaboration} {Virgo, LIGO
  Scientific}),\ }\href {\doibase 10.1103/PhysRevLett.119.161101} {\bibfield
  {journal} {\bibinfo  {journal} {Phys. Rev. Lett.}\ }\textbf {\bibinfo
  {volume} {119}},\ \bibinfo {pages} {161101} (\bibinfo {year}
  {2017}{\natexlab{c}})},\ \Eprint {http://arxiv.org/abs/1710.05832}
  {arXiv:1710.05832 [gr-qc]} \BibitemShut {NoStop}%
\bibitem [{\citenamefont {Abbott}\ \emph
  {et~al.}(2017{\natexlab{d}})\citenamefont {Abbott} \emph
  {et~al.}}]{Abbott:2017gyy}%
  \BibitemOpen
  \bibfield  {author} {\bibinfo {author} {\bibfnamefont {B.~P.}\ \bibnamefont
  {Abbott}} \emph {et~al.} (\bibinfo {collaboration} {Virgo, LIGO
  Scientific}),\ }\href {\doibase 10.3847/2041-8213/aa9f0c} {\bibfield
  {journal} {\bibinfo  {journal} {Astrophys. J.}\ }\textbf {\bibinfo {volume}
  {851}},\ \bibinfo {pages} {L35} (\bibinfo {year} {2017}{\natexlab{d}})},\
  \Eprint {http://arxiv.org/abs/1711.05578} {arXiv:1711.05578 [astro-ph.HE]}
  \BibitemShut {NoStop}%
\bibitem [{\citenamefont {Abbott}\ \emph {et~al.}(2019)\citenamefont {Abbott}
  \emph {et~al.}}]{LIGOScientific:2018mvr}%
  \BibitemOpen
  \bibfield  {author} {\bibinfo {author} {\bibfnamefont {B.~P.}\ \bibnamefont
  {Abbott}} \emph {et~al.} (\bibinfo {collaboration} {LIGO Scientific,
  Virgo}),\ }\href {\doibase 10.1103/PhysRevX.9.031040} {\bibfield  {journal}
  {\bibinfo  {journal} {Phys. Rev. X}\ }\textbf {\bibinfo {volume} {9}},\
  \bibinfo {pages} {031040} (\bibinfo {year} {2019})},\ \Eprint
  {http://arxiv.org/abs/1811.12907} {arXiv:1811.12907 [astro-ph.HE]}
  \BibitemShut {NoStop}%
\bibitem [{\citenamefont {Abbott}\ \emph {et~al.}(2020)\citenamefont {Abbott}
  \emph {et~al.}}]{Abbott:2020uma}%
  \BibitemOpen
  \bibfield  {author} {\bibinfo {author} {\bibfnamefont {B.~P.}\ \bibnamefont
  {Abbott}} \emph {et~al.} (\bibinfo {collaboration} {LIGO Scientific,
  Virgo}),\ }\href@noop {} {\  (\bibinfo {year} {2020})},\ \Eprint
  {http://arxiv.org/abs/2001.01761} {arXiv:2001.01761 [astro-ph.HE]}
  \BibitemShut {NoStop}%
\bibitem [{\citenamefont {de~Paula}\ \emph {et~al.}(2004)\citenamefont
  {de~Paula}, \citenamefont {Miranda},\ and\ \citenamefont
  {Marinho}}]{dePaula:2004bc}%
  \BibitemOpen
  \bibfield  {author} {\bibinfo {author} {\bibfnamefont {W.~L.~S.}\
  \bibnamefont {de~Paula}}, \bibinfo {author} {\bibfnamefont {O.~D.}\
  \bibnamefont {Miranda}}, \ and\ \bibinfo {author} {\bibfnamefont {R.~M.}\
  \bibnamefont {Marinho}},\ }\href {\doibase 10.1088/0264-9381/21/19/008}
  {\bibfield  {journal} {\bibinfo  {journal} {Class. Quant. Grav.}\ }\textbf
  {\bibinfo {volume} {21}},\ \bibinfo {pages} {4595} (\bibinfo {year}
  {2004})},\ \Eprint {http://arxiv.org/abs/gr-qc/0409041} {arXiv:gr-qc/0409041
  [gr-qc]} \BibitemShut {NoStop}%
\bibitem [{\citenamefont {Nishizawa}\ \emph {et~al.}(2009)\citenamefont
  {Nishizawa}, \citenamefont {Taruya}, \citenamefont {Hayama}, \citenamefont
  {Kawamura},\ and\ \citenamefont {Sakagami}}]{Nishizawa:2009bf}%
  \BibitemOpen
  \bibfield  {author} {\bibinfo {author} {\bibfnamefont {A.}~\bibnamefont
  {Nishizawa}}, \bibinfo {author} {\bibfnamefont {A.}~\bibnamefont {Taruya}},
  \bibinfo {author} {\bibfnamefont {K.}~\bibnamefont {Hayama}}, \bibinfo
  {author} {\bibfnamefont {S.}~\bibnamefont {Kawamura}}, \ and\ \bibinfo
  {author} {\bibfnamefont {M.-a.}\ \bibnamefont {Sakagami}},\ }\href {\doibase
  10.1103/PhysRevD.79.082002} {\bibfield  {journal} {\bibinfo  {journal} {Phys.
  Rev.}\ }\textbf {\bibinfo {volume} {D79}},\ \bibinfo {pages} {082002}
  (\bibinfo {year} {2009})},\ \Eprint {http://arxiv.org/abs/0903.0528}
  {arXiv:0903.0528 [astro-ph.CO]} \BibitemShut {NoStop}%
\bibitem [{\citenamefont {Will}(2014)}]{Will:2014kxa}%
  \BibitemOpen
  \bibfield  {author} {\bibinfo {author} {\bibfnamefont {C.~M.}\ \bibnamefont
  {Will}},\ }\href {\doibase 10.12942/lrr-2014-4} {\bibfield  {journal}
  {\bibinfo  {journal} {Living Rev. Rel.}\ }\textbf {\bibinfo {volume} {17}},\
  \bibinfo {pages} {4} (\bibinfo {year} {2014})},\ \Eprint
  {http://arxiv.org/abs/1403.7377} {arXiv:1403.7377 [gr-qc]} \BibitemShut
  {NoStop}%
\bibitem [{\citenamefont {Hou}\ \emph {et~al.}(2018)\citenamefont {Hou},
  \citenamefont {Gong},\ and\ \citenamefont {Liu}}]{Hou:2017bqj}%
  \BibitemOpen
  \bibfield  {author} {\bibinfo {author} {\bibfnamefont {S.}~\bibnamefont
  {Hou}}, \bibinfo {author} {\bibfnamefont {Y.}~\bibnamefont {Gong}}, \ and\
  \bibinfo {author} {\bibfnamefont {Y.}~\bibnamefont {Liu}},\ }\href {\doibase
  10.1140/epjc/s10052-018-5869-y} {\bibfield  {journal} {\bibinfo  {journal}
  {Eur. Phys. J. C}\ }\textbf {\bibinfo {volume} {78}},\ \bibinfo {pages} {378}
  (\bibinfo {year} {2018})},\ \Eprint {http://arxiv.org/abs/1704.01899}
  {arXiv:1704.01899 [gr-qc]} \BibitemShut {NoStop}%
\bibitem [{\citenamefont {{Hou}}\ and\ \citenamefont
  {{Gong}}(2018)}]{Hou:2018djz}%
  \BibitemOpen
  \bibfield  {author} {\bibinfo {author} {\bibfnamefont {S.}~\bibnamefont
  {{Hou}}}\ and\ \bibinfo {author} {\bibfnamefont {Y.}~\bibnamefont {{Gong}}},\
  }\href {\doibase 10.3390/universe4080084} {\bibfield  {journal} {\bibinfo
  {journal} {Universe}\ }\textbf {\bibinfo {volume} {4}},\ \bibinfo {pages}
  {84} (\bibinfo {year} {2018})},\ \Eprint {http://arxiv.org/abs/1806.02564}
  {arXiv:1806.02564 [gr-qc]} \BibitemShut {NoStop}%
\bibitem [{\citenamefont {{Gong}}\ and\ \citenamefont
  {{Hou}}(2018)}]{Gong:2018ybk}%
  \BibitemOpen
  \bibfield  {author} {\bibinfo {author} {\bibfnamefont {Y.}~\bibnamefont
  {{Gong}}}\ and\ \bibinfo {author} {\bibfnamefont {S.}~\bibnamefont {{Hou}}},\
  }\href {\doibase 10.3390/universe4080085} {\bibfield  {journal} {\bibinfo
  {journal} {Universe}\ }\textbf {\bibinfo {volume} {4}},\ \bibinfo {pages}
  {85} (\bibinfo {year} {2018})},\ \Eprint {http://arxiv.org/abs/1806.04027}
  {arXiv:1806.04027 [gr-qc]} \BibitemShut {NoStop}%
\bibitem [{\citenamefont {Gong}\ \emph {et~al.}(2018)\citenamefont {Gong},
  \citenamefont {Hou}, \citenamefont {Liang},\ and\ \citenamefont
  {Papantonopoulos}}]{Gong:2018cgj}%
  \BibitemOpen
  \bibfield  {author} {\bibinfo {author} {\bibfnamefont {Y.}~\bibnamefont
  {Gong}}, \bibinfo {author} {\bibfnamefont {S.}~\bibnamefont {Hou}}, \bibinfo
  {author} {\bibfnamefont {D.}~\bibnamefont {Liang}}, \ and\ \bibinfo {author}
  {\bibfnamefont {E.}~\bibnamefont {Papantonopoulos}},\ }\href {\doibase
  10.1103/PhysRevD.97.084040} {\bibfield  {journal} {\bibinfo  {journal} {Phys.
  Rev. D}\ }\textbf {\bibinfo {volume} {97}},\ \bibinfo {pages} {084040}
  (\bibinfo {year} {2018})},\ \Eprint {http://arxiv.org/abs/1801.03382}
  {arXiv:1801.03382 [gr-qc]} \BibitemShut {NoStop}%
\bibitem [{\citenamefont {Hou}\ and\ \citenamefont {Gong}(2019)}]{Hou:2018mey}%
  \BibitemOpen
  \bibfield  {author} {\bibinfo {author} {\bibfnamefont {S.}~\bibnamefont
  {Hou}}\ and\ \bibinfo {author} {\bibfnamefont {Y.}~\bibnamefont {Gong}},\
  }\href {\doibase 10.1140/epjc/s10052-019-6684-9} {\bibfield  {journal}
  {\bibinfo  {journal} {Eur. Phys. J. C}\ }\textbf {\bibinfo {volume} {79}},\
  \bibinfo {pages} {197} (\bibinfo {year} {2019})},\ \Eprint
  {http://arxiv.org/abs/1810.00630} {arXiv:1810.00630 [gr-qc]} \BibitemShut
  {NoStop}%
\bibitem [{\citenamefont {Soudi}\ \emph {et~al.}(2019)\citenamefont {Soudi},
  \citenamefont {Farrugia}, \citenamefont {Gakis}, \citenamefont {Levi~Said},\
  and\ \citenamefont {Saridakis}}]{Soudi:2018dhv}%
  \BibitemOpen
  \bibfield  {author} {\bibinfo {author} {\bibfnamefont {I.}~\bibnamefont
  {Soudi}}, \bibinfo {author} {\bibfnamefont {G.}~\bibnamefont {Farrugia}},
  \bibinfo {author} {\bibfnamefont {V.}~\bibnamefont {Gakis}}, \bibinfo
  {author} {\bibfnamefont {J.}~\bibnamefont {Levi~Said}}, \ and\ \bibinfo
  {author} {\bibfnamefont {E.~N.}\ \bibnamefont {Saridakis}},\ }\href {\doibase
  10.1103/PhysRevD.100.044008} {\bibfield  {journal} {\bibinfo  {journal}
  {Phys. Rev.}\ }\textbf {\bibinfo {volume} {D100}},\ \bibinfo {pages} {044008}
  (\bibinfo {year} {2019})},\ \Eprint {http://arxiv.org/abs/1810.08220}
  {arXiv:1810.08220 [gr-qc]} \BibitemShut {NoStop}%
\bibitem [{\citenamefont {Hohmann}\ \emph {et~al.}(2019)\citenamefont
  {Hohmann}, \citenamefont {J\"arv}, \citenamefont {Kr\v{s}\v{s}\'ak},\ and\
  \citenamefont {Pfeifer}}]{Hohmann:2019nat}%
  \BibitemOpen
  \bibfield  {author} {\bibinfo {author} {\bibfnamefont {M.}~\bibnamefont
  {Hohmann}}, \bibinfo {author} {\bibfnamefont {L.}~\bibnamefont {J\"arv}},
  \bibinfo {author} {\bibfnamefont {M.}~\bibnamefont {Kr\v{s}\v{s}\'ak}}, \
  and\ \bibinfo {author} {\bibfnamefont {C.}~\bibnamefont {Pfeifer}},\ }\href
  {\doibase 10.1103/PhysRevD.100.084002} {\bibfield  {journal} {\bibinfo
  {journal} {Phys. Rev.}\ }\textbf {\bibinfo {volume} {D100}},\ \bibinfo
  {pages} {084002} (\bibinfo {year} {2019})},\ \Eprint
  {http://arxiv.org/abs/1901.05472} {arXiv:1901.05472 [gr-qc]} \BibitemShut
  {NoStop}%
\bibitem [{\citenamefont {Liu}\ \emph {et~al.}(2019{\natexlab{a}})\citenamefont
  {Liu}, \citenamefont {Qian}, \citenamefont {Gong},\ and\ \citenamefont
  {Wang}}]{Liu:2019cxm}%
  \BibitemOpen
  \bibfield  {author} {\bibinfo {author} {\bibfnamefont {Y.}~\bibnamefont
  {Liu}}, \bibinfo {author} {\bibfnamefont {W.-L.}\ \bibnamefont {Qian}},
  \bibinfo {author} {\bibfnamefont {Y.}~\bibnamefont {Gong}}, \ and\ \bibinfo
  {author} {\bibfnamefont {B.}~\bibnamefont {Wang}},\ }\href@noop {} {\
  (\bibinfo {year} {2019}{\natexlab{a}})},\ \Eprint
  {http://arxiv.org/abs/1912.01420} {arXiv:1912.01420 [gr-qc]} \BibitemShut
  {NoStop}%
\bibitem [{\citenamefont {Misner}\ \emph {et~al.}(1973)\citenamefont {Misner},
  \citenamefont {Thorne},\ and\ \citenamefont {Wheeler}}]{Misner:1974qy}%
  \BibitemOpen
  \bibfield  {author} {\bibinfo {author} {\bibfnamefont {C.~W.}\ \bibnamefont
  {Misner}}, \bibinfo {author} {\bibfnamefont {K.~S.}\ \bibnamefont {Thorne}},
  \ and\ \bibinfo {author} {\bibfnamefont {J.~A.}\ \bibnamefont {Wheeler}},\
  }\href@noop {} {\emph {\bibinfo {title} {{Gravitation}}}}\ (\bibinfo
  {publisher} {W. H. Freeman},\ \bibinfo {address} {San Francisco},\ \bibinfo
  {year} {1973})\BibitemShut {NoStop}%
\bibitem [{\citenamefont {{Schneider}}\ \emph {et~al.}(1992)\citenamefont
  {{Schneider}}, \citenamefont {{Ehlers}},\ and\ \citenamefont
  {{Falco}}}]{gravlens1992}%
  \BibitemOpen
  \bibfield  {author} {\bibinfo {author} {\bibfnamefont {P.}~\bibnamefont
  {{Schneider}}}, \bibinfo {author} {\bibfnamefont {J.}~\bibnamefont
  {{Ehlers}}}, \ and\ \bibinfo {author} {\bibfnamefont {E.~E.}\ \bibnamefont
  {{Falco}}},\ }\href {\doibase 10.1007/978-3-662-03758-4} {\emph {\bibinfo
  {title} {Gravitational Lenses}}}\ (\bibinfo  {publisher} {Springer, Berlin,
  Heidelberg},\ \bibinfo {year} {1992})\ p.\ \bibinfo {pages} {560}\BibitemShut
  {NoStop}%
\bibitem [{\citenamefont {Lawrence}(1971{\natexlab{a}})}]{Lawrence1971nc}%
  \BibitemOpen
  \bibfield  {author} {\bibinfo {author} {\bibfnamefont {J.~K.}\ \bibnamefont
  {Lawrence}},\ }\href {\doibase 10.1007/BF02735388} {\bibfield  {journal}
  {\bibinfo  {journal} {Il Nuovo Cimento B (1971-1996)}\ }\textbf {\bibinfo
  {volume} {6}},\ \bibinfo {pages} {225} (\bibinfo {year}
  {1971}{\natexlab{a}})}\BibitemShut {NoStop}%
\bibitem [{\citenamefont {Lawrence}(1971{\natexlab{b}})}]{Lawrence:1971hx}%
  \BibitemOpen
  \bibfield  {author} {\bibinfo {author} {\bibfnamefont {J.~K.}\ \bibnamefont
  {Lawrence}},\ }\href {\doibase 10.1103/PhysRevD.3.3239} {\bibfield  {journal}
  {\bibinfo  {journal} {Phys. Rev. D}\ }\textbf {\bibinfo {volume} {3}},\
  \bibinfo {pages} {3239} (\bibinfo {year} {1971}{\natexlab{b}})}\BibitemShut
  {NoStop}%
\bibitem [{\citenamefont {Ng}\ \emph {et~al.}(2018)\citenamefont {Ng},
  \citenamefont {Wong}, \citenamefont {Broadhurst},\ and\ \citenamefont
  {Li}}]{Ng:2017yiu}%
  \BibitemOpen
  \bibfield  {author} {\bibinfo {author} {\bibfnamefont {K.~K.~Y.}\
  \bibnamefont {Ng}}, \bibinfo {author} {\bibfnamefont {K.~W.~K.}\ \bibnamefont
  {Wong}}, \bibinfo {author} {\bibfnamefont {T.}~\bibnamefont {Broadhurst}}, \
  and\ \bibinfo {author} {\bibfnamefont {T.~G.~F.}\ \bibnamefont {Li}},\ }\href
  {\doibase 10.1103/PhysRevD.97.023012} {\bibfield  {journal} {\bibinfo
  {journal} {Phys. Rev. D}\ }\textbf {\bibinfo {volume} {97}},\ \bibinfo
  {pages} {023012} (\bibinfo {year} {2018})},\ \Eprint
  {http://arxiv.org/abs/1703.06319} {arXiv:1703.06319 [astro-ph.CO]}
  \BibitemShut {NoStop}%
\bibitem [{\citenamefont {Dai}\ \emph {et~al.}(2017)\citenamefont {Dai},
  \citenamefont {Venumadhav},\ and\ \citenamefont {Sigurdson}}]{Dai:2016igl}%
  \BibitemOpen
  \bibfield  {author} {\bibinfo {author} {\bibfnamefont {L.}~\bibnamefont
  {Dai}}, \bibinfo {author} {\bibfnamefont {T.}~\bibnamefont {Venumadhav}}, \
  and\ \bibinfo {author} {\bibfnamefont {K.}~\bibnamefont {Sigurdson}},\ }\href
  {\doibase 10.1103/PhysRevD.95.044011} {\bibfield  {journal} {\bibinfo
  {journal} {Phys. Rev. D}\ }\textbf {\bibinfo {volume} {95}},\ \bibinfo
  {pages} {044011} (\bibinfo {year} {2017})},\ \Eprint
  {http://arxiv.org/abs/1605.09398} {arXiv:1605.09398 [astro-ph.CO]}
  \BibitemShut {NoStop}%
\bibitem [{\citenamefont {Nakamura}(1998)}]{Nakamura:1997sw}%
  \BibitemOpen
  \bibfield  {author} {\bibinfo {author} {\bibfnamefont {T.~T.}\ \bibnamefont
  {Nakamura}},\ }\href {\doibase 10.1103/PhysRevLett.80.1138} {\bibfield
  {journal} {\bibinfo  {journal} {Phys. Rev. Lett.}\ }\textbf {\bibinfo
  {volume} {80}},\ \bibinfo {pages} {1138} (\bibinfo {year}
  {1998})}\BibitemShut {NoStop}%
\bibitem [{\citenamefont {{Nakamura}}\ and\ \citenamefont
  {{Deguchi}}(1999)}]{Nakamura1999wo}%
  \BibitemOpen
  \bibfield  {author} {\bibinfo {author} {\bibfnamefont {T.~T.}\ \bibnamefont
  {{Nakamura}}}\ and\ \bibinfo {author} {\bibfnamefont {S.}~\bibnamefont
  {{Deguchi}}},\ }\href {\doibase 10.1143/PTPS.133.137} {\bibfield  {journal}
  {\bibinfo  {journal} {Progress of Theoretical Physics Supplement}\ }\textbf
  {\bibinfo {volume} {133}},\ \bibinfo {pages} {137} (\bibinfo {year}
  {1999})}\BibitemShut {NoStop}%
\bibitem [{\citenamefont {Takahashi}\ and\ \citenamefont
  {Nakamura}(2003)}]{Takahashi:2003ix}%
  \BibitemOpen
  \bibfield  {author} {\bibinfo {author} {\bibfnamefont {R.}~\bibnamefont
  {Takahashi}}\ and\ \bibinfo {author} {\bibfnamefont {T.}~\bibnamefont
  {Nakamura}},\ }\href {\doibase 10.1086/377430} {\bibfield  {journal}
  {\bibinfo  {journal} {Astrophys. J.}\ }\textbf {\bibinfo {volume} {595}},\
  \bibinfo {pages} {1039} (\bibinfo {year} {2003})},\ \Eprint
  {http://arxiv.org/abs/astro-ph/0305055} {arXiv:astro-ph/0305055 [astro-ph]}
  \BibitemShut {NoStop}%
\bibitem [{\citenamefont {Takahashi}(2017)}]{Takahashi:2016jom}%
  \BibitemOpen
  \bibfield  {author} {\bibinfo {author} {\bibfnamefont {R.}~\bibnamefont
  {Takahashi}},\ }\href {\doibase 10.3847/1538-4357/835/1/103} {\bibfield
  {journal} {\bibinfo  {journal} {Astrophys. J.}\ }\textbf {\bibinfo {volume}
  {835}},\ \bibinfo {pages} {103} (\bibinfo {year} {2017})},\ \Eprint
  {http://arxiv.org/abs/1606.00458} {arXiv:1606.00458 [astro-ph.CO]}
  \BibitemShut {NoStop}%
\bibitem [{\citenamefont {Arnaud-Varvella}\ \emph {et~al.}(2004)\citenamefont
  {Arnaud-Varvella}, \citenamefont {Angonin},\ and\ \citenamefont
  {Tourrenc}}]{ArnaudVarvella:2003va}%
  \BibitemOpen
  \bibfield  {author} {\bibinfo {author} {\bibfnamefont {M.}~\bibnamefont
  {Arnaud-Varvella}}, \bibinfo {author} {\bibfnamefont {M.~C.}\ \bibnamefont
  {Angonin}}, \ and\ \bibinfo {author} {\bibfnamefont {P.}~\bibnamefont
  {Tourrenc}},\ }\href {\doibase 10.1023/B:GERG.0000018085.46454.ff} {\bibfield
   {journal} {\bibinfo  {journal} {Gen. Rel. Grav.}\ }\textbf {\bibinfo
  {volume} {36}},\ \bibinfo {pages} {983} (\bibinfo {year} {2004})},\ \Eprint
  {http://arxiv.org/abs/gr-qc/0312028} {arXiv:gr-qc/0312028 [gr-qc]}
  \BibitemShut {NoStop}%
\bibitem [{\citenamefont {Meena}\ and\ \citenamefont
  {Bagla}(2019)}]{Meena:2019ate}%
  \BibitemOpen
  \bibfield  {author} {\bibinfo {author} {\bibfnamefont {A.~K.}\ \bibnamefont
  {Meena}}\ and\ \bibinfo {author} {\bibfnamefont {J.~S.}\ \bibnamefont
  {Bagla}},\ }\href {\doibase 10.1093/mnras/stz3509} {\bibfield  {journal}
  {\bibinfo  {journal} {Mon. Not. Roy. Astron. Soc.}\ } (\bibinfo {year}
  {2019}),\ 10.1093/mnras/stz3509},\ \Eprint {http://arxiv.org/abs/1903.11809}
  {arXiv:1903.11809 [astro-ph.CO]} \BibitemShut {NoStop}%
\bibitem [{\citenamefont {Kawamura}\ \emph {et~al.}(2011)\citenamefont
  {Kawamura} \emph {et~al.}}]{Kawamura:2011zz}%
  \BibitemOpen
  \bibfield  {author} {\bibinfo {author} {\bibfnamefont {S.}~\bibnamefont
  {Kawamura}} \emph {et~al.},\ }\bibfield  {booktitle} {\emph {\bibinfo
  {booktitle} {{Laser interferometer space antenna. Proceedings, 8th
  International LISA Symposium, Stanford, USA, June 28-July 2, 2010}}},\ }\href
  {\doibase 10.1088/0264-9381/28/9/094011} {\bibfield  {journal} {\bibinfo
  {journal} {Class. Quant. Grav.}\ }\textbf {\bibinfo {volume} {28}},\ \bibinfo
  {pages} {094011} (\bibinfo {year} {2011})}\BibitemShut {NoStop}%
\bibitem [{\citenamefont {Liao}\ \emph {et~al.}(2019)\citenamefont {Liao},
  \citenamefont {Biesiada},\ and\ \citenamefont {Fan}}]{Liao:2019aqq}%
  \BibitemOpen
  \bibfield  {author} {\bibinfo {author} {\bibfnamefont {K.}~\bibnamefont
  {Liao}}, \bibinfo {author} {\bibfnamefont {M.}~\bibnamefont {Biesiada}}, \
  and\ \bibinfo {author} {\bibfnamefont {X.-L.}\ \bibnamefont {Fan}},\ }\href
  {\doibase 10.3847/1538-4357/ab1087} {\bibfield  {journal} {\bibinfo
  {journal} {Astrophys. J.}\ }\textbf {\bibinfo {volume} {875}},\ \bibinfo
  {pages} {139} (\bibinfo {year} {2019})},\ \Eprint
  {http://arxiv.org/abs/1903.06612} {arXiv:1903.06612 [gr-qc]} \BibitemShut
  {NoStop}%
\bibitem [{\citenamefont {Hou}\ \emph {et~al.}(2019)\citenamefont {Hou},
  \citenamefont {Fan},\ and\ \citenamefont {Zhu}}]{Hou:2019wdg}%
  \BibitemOpen
  \bibfield  {author} {\bibinfo {author} {\bibfnamefont {S.}~\bibnamefont
  {Hou}}, \bibinfo {author} {\bibfnamefont {X.-L.}\ \bibnamefont {Fan}}, \ and\
  \bibinfo {author} {\bibfnamefont {Z.-H.}\ \bibnamefont {Zhu}},\ }\href
  {\doibase 10.1103/PhysRevD.100.064028} {\bibfield  {journal} {\bibinfo
  {journal} {Phys. Rev. D}\ }\textbf {\bibinfo {volume} {100}},\ \bibinfo
  {pages} {064028} (\bibinfo {year} {2019})},\ \Eprint
  {http://arxiv.org/abs/1907.07486} {arXiv:1907.07486 [gr-qc]} \BibitemShut
  {NoStop}%
\bibitem [{\citenamefont {Schilling}(1997)}]{Schilling:1997id}%
  \BibitemOpen
  \bibfield  {author} {\bibinfo {author} {\bibfnamefont {R.}~\bibnamefont
  {Schilling}},\ }\bibfield  {booktitle} {\emph {\bibinfo {booktitle} {{1st
  International LISA Symposium on Gravitational Waves Oxfordshire, England,
  July 9-12, 1996}}},\ }\href {\doibase 10.1088/0264-9381/14/6/020} {\bibfield
  {journal} {\bibinfo  {journal} {Class. Quant. Grav.}\ }\textbf {\bibinfo
  {volume} {14}},\ \bibinfo {pages} {1513} (\bibinfo {year}
  {1997})}\BibitemShut {NoStop}%
\bibitem [{\citenamefont {Giampieri}(1997)}]{Giampieri:1997kv}%
  \BibitemOpen
  \bibfield  {author} {\bibinfo {author} {\bibfnamefont {G.}~\bibnamefont
  {Giampieri}},\ }\href {\doibase 10.1093/mnras/289.1.185} {\bibfield
  {journal} {\bibinfo  {journal} {Mon. Not. Roy. Astron. Soc.}\ }\textbf
  {\bibinfo {volume} {289}},\ \bibinfo {pages} {185} (\bibinfo {year}
  {1997})},\ \Eprint {http://arxiv.org/abs/gr-qc/9704069} {arXiv:gr-qc/9704069
  [gr-qc]} \BibitemShut {NoStop}%
\bibitem [{\citenamefont {Liang}\ \emph {et~al.}(2019)\citenamefont {Liang},
  \citenamefont {Gong}, \citenamefont {Weinstein}, \citenamefont {Zhang},\ and\
  \citenamefont {Zhang}}]{Liang:2019pry}%
  \BibitemOpen
  \bibfield  {author} {\bibinfo {author} {\bibfnamefont {D.}~\bibnamefont
  {Liang}}, \bibinfo {author} {\bibfnamefont {Y.}~\bibnamefont {Gong}},
  \bibinfo {author} {\bibfnamefont {A.~J.}\ \bibnamefont {Weinstein}}, \bibinfo
  {author} {\bibfnamefont {C.}~\bibnamefont {Zhang}}, \ and\ \bibinfo {author}
  {\bibfnamefont {C.}~\bibnamefont {Zhang}},\ }\href {\doibase
  10.1103/PhysRevD.99.104027} {\bibfield  {journal} {\bibinfo  {journal} {Phys.
  Rev. D}\ }\textbf {\bibinfo {volume} {99}},\ \bibinfo {pages} {104027}
  (\bibinfo {year} {2019})},\ \Eprint {http://arxiv.org/abs/1901.09624}
  {arXiv:1901.09624 [gr-qc]} \BibitemShut {NoStop}%
\bibitem [{\citenamefont {Diego}\ \emph {et~al.}(2019)\citenamefont {Diego},
  \citenamefont {Hannuksela}, \citenamefont {Kelly}, \citenamefont
  {Broadhurst}, \citenamefont {Kim}, \citenamefont {Li}, \citenamefont
  {Smoot},\ and\ \citenamefont {Pagano}}]{Diego:2019lcd}%
  \BibitemOpen
  \bibfield  {author} {\bibinfo {author} {\bibfnamefont {J.~M.}\ \bibnamefont
  {Diego}}, \bibinfo {author} {\bibfnamefont {O.~A.}\ \bibnamefont
  {Hannuksela}}, \bibinfo {author} {\bibfnamefont {P.~L.}\ \bibnamefont
  {Kelly}}, \bibinfo {author} {\bibfnamefont {T.}~\bibnamefont {Broadhurst}},
  \bibinfo {author} {\bibfnamefont {K.}~\bibnamefont {Kim}}, \bibinfo {author}
  {\bibfnamefont {T.~G.~F.}\ \bibnamefont {Li}}, \bibinfo {author}
  {\bibfnamefont {G.~F.}\ \bibnamefont {Smoot}}, \ and\ \bibinfo {author}
  {\bibfnamefont {G.}~\bibnamefont {Pagano}},\ }\href {\doibase
  10.1051/0004-6361/201935490} {\bibfield  {journal} {\bibinfo  {journal}
  {Astron. Astrophys.}\ }\textbf {\bibinfo {volume} {627}},\ \bibinfo {pages}
  {A130} (\bibinfo {year} {2019})},\ \Eprint {http://arxiv.org/abs/1903.04513}
  {arXiv:1903.04513 [astro-ph.CO]} \BibitemShut {NoStop}%
\bibitem [{Note1()}]{Note1}%
  \BibitemOpen
  \bibinfo {note} {According to Ref.~\cite {Diego:2019lcd}, in the saturation
  regime, the caustics corresponding to the microlenses contained in the
  galaxies overlap in the source plane. So GWs emitted by sources near or
  moving across this regime will be microlensed. Refer to Ref.~\cite
  {Diego:2017drh}, too.}\BibitemShut {Stop}%
\bibitem [{\citenamefont {Schutz}(1986)}]{Schutz:1986gp}%
  \BibitemOpen
  \bibfield  {author} {\bibinfo {author} {\bibfnamefont {B.~F.}\ \bibnamefont
  {Schutz}},\ }\href {\doibase 10.1038/323310a0} {\bibfield  {journal}
  {\bibinfo  {journal} {Nature}\ }\textbf {\bibinfo {volume} {323}},\ \bibinfo
  {pages} {310} (\bibinfo {year} {1986})}\BibitemShut {NoStop}%
\bibitem [{\citenamefont {Holz}\ and\ \citenamefont
  {Hughes}(2005)}]{Holz:2005df}%
  \BibitemOpen
  \bibfield  {author} {\bibinfo {author} {\bibfnamefont {D.~E.}\ \bibnamefont
  {Holz}}\ and\ \bibinfo {author} {\bibfnamefont {S.~A.}\ \bibnamefont
  {Hughes}},\ }\href {\doibase 10.1086/431341} {\bibfield  {journal} {\bibinfo
  {journal} {Astrophys. J.}\ }\textbf {\bibinfo {volume} {629}},\ \bibinfo
  {pages} {15} (\bibinfo {year} {2005})},\ \Eprint
  {http://arxiv.org/abs/astro-ph/0504616} {arXiv:astro-ph/0504616 [astro-ph]}
  \BibitemShut {NoStop}%
\bibitem [{\citenamefont {Shapiro}\ \emph {et~al.}(2010)\citenamefont
  {Shapiro}, \citenamefont {Bacon}, \citenamefont {Hendry},\ and\ \citenamefont
  {Hoyle}}]{Shapiro:2009sr}%
  \BibitemOpen
  \bibfield  {author} {\bibinfo {author} {\bibfnamefont {C.}~\bibnamefont
  {Shapiro}}, \bibinfo {author} {\bibfnamefont {D.}~\bibnamefont {Bacon}},
  \bibinfo {author} {\bibfnamefont {M.}~\bibnamefont {Hendry}}, \ and\ \bibinfo
  {author} {\bibfnamefont {B.}~\bibnamefont {Hoyle}},\ }\href {\doibase
  10.1111/j.1365-2966.2010.16317.x} {\bibfield  {journal} {\bibinfo  {journal}
  {Mon. Not. Roy. Astron. Soc.}\ }\textbf {\bibinfo {volume} {404}},\ \bibinfo
  {pages} {858} (\bibinfo {year} {2010})},\ \Eprint
  {http://arxiv.org/abs/0907.3635} {arXiv:0907.3635 [astro-ph.CO]} \BibitemShut
  {NoStop}%
\bibitem [{\citenamefont {Hilbert}\ \emph {et~al.}(2011)\citenamefont
  {Hilbert}, \citenamefont {Gair},\ and\ \citenamefont
  {King}}]{Hilbert:2010am}%
  \BibitemOpen
  \bibfield  {author} {\bibinfo {author} {\bibfnamefont {S.}~\bibnamefont
  {Hilbert}}, \bibinfo {author} {\bibfnamefont {J.~R.}\ \bibnamefont {Gair}}, \
  and\ \bibinfo {author} {\bibfnamefont {L.~J.}\ \bibnamefont {King}},\ }\href
  {\doibase 10.1111/j.1365-2966.2010.17963.x} {\bibfield  {journal} {\bibinfo
  {journal} {Mon. Not. Roy. Astron. Soc.}\ }\textbf {\bibinfo {volume} {412}},\
  \bibinfo {pages} {1023} (\bibinfo {year} {2011})},\ \Eprint
  {http://arxiv.org/abs/1007.2468} {arXiv:1007.2468 [astro-ph.CO]} \BibitemShut
  {NoStop}%
\bibitem [{\citenamefont {Yamamoto}(2005)}]{Yamamoto:2005ea}%
  \BibitemOpen
  \bibfield  {author} {\bibinfo {author} {\bibfnamefont {K.}~\bibnamefont
  {Yamamoto}},\ }\href {\doibase 10.1103/PhysRevD.71.101301} {\bibfield
  {journal} {\bibinfo  {journal} {Phys. Rev. D}\ }\textbf {\bibinfo {volume}
  {71}},\ \bibinfo {pages} {101301(R)} (\bibinfo {year} {2005})},\ \Eprint
  {http://arxiv.org/abs/astro-ph/0505116} {arXiv:astro-ph/0505116 [astro-ph]}
  \BibitemShut {NoStop}%
\bibitem [{\citenamefont {Cutler}\ and\ \citenamefont
  {Holz}(2009)}]{Cutler:2009qv}%
  \BibitemOpen
  \bibfield  {author} {\bibinfo {author} {\bibfnamefont {C.}~\bibnamefont
  {Cutler}}\ and\ \bibinfo {author} {\bibfnamefont {D.~E.}\ \bibnamefont
  {Holz}},\ }\href {\doibase 10.1103/PhysRevD.80.104009} {\bibfield  {journal}
  {\bibinfo  {journal} {Phys. Rev. D}\ }\textbf {\bibinfo {volume} {80}},\
  \bibinfo {pages} {104009} (\bibinfo {year} {2009})},\ \Eprint
  {http://arxiv.org/abs/0906.3752} {arXiv:0906.3752 [astro-ph.CO]} \BibitemShut
  {NoStop}%
\bibitem [{\citenamefont {Camera}\ and\ \citenamefont
  {Nishizawa}(2013)}]{Camera:2013xfa}%
  \BibitemOpen
  \bibfield  {author} {\bibinfo {author} {\bibfnamefont {S.}~\bibnamefont
  {Camera}}\ and\ \bibinfo {author} {\bibfnamefont {A.}~\bibnamefont
  {Nishizawa}},\ }\href {\doibase 10.1103/PhysRevLett.110.151103} {\bibfield
  {journal} {\bibinfo  {journal} {Phys. Rev. Lett.}\ }\textbf {\bibinfo
  {volume} {110}},\ \bibinfo {pages} {151103} (\bibinfo {year} {2013})},\
  \Eprint {http://arxiv.org/abs/1303.5446} {arXiv:1303.5446 [astro-ph.CO]}
  \BibitemShut {NoStop}%
\bibitem [{\citenamefont {Congedo}\ and\ \citenamefont
  {Taylor}(2019)}]{Congedo:2018wfn}%
  \BibitemOpen
  \bibfield  {author} {\bibinfo {author} {\bibfnamefont {G.}~\bibnamefont
  {Congedo}}\ and\ \bibinfo {author} {\bibfnamefont {A.}~\bibnamefont
  {Taylor}},\ }\href {\doibase 10.1103/PhysRevD.99.083526} {\bibfield
  {journal} {\bibinfo  {journal} {Phys. Rev. D}\ }\textbf {\bibinfo {volume}
  {99}},\ \bibinfo {pages} {083526} (\bibinfo {year} {2019})},\ \Eprint
  {http://arxiv.org/abs/1812.02730} {arXiv:1812.02730 [astro-ph.CO]}
  \BibitemShut {NoStop}%
\bibitem [{\citenamefont {Jung}\ and\ \citenamefont
  {Shin}(2019)}]{Jung:2017flg}%
  \BibitemOpen
  \bibfield  {author} {\bibinfo {author} {\bibfnamefont {S.}~\bibnamefont
  {Jung}}\ and\ \bibinfo {author} {\bibfnamefont {C.~S.}\ \bibnamefont
  {Shin}},\ }\href {\doibase 10.1103/PhysRevLett.122.041103} {\bibfield
  {journal} {\bibinfo  {journal} {Phys. Rev. Lett.}\ }\textbf {\bibinfo
  {volume} {122}},\ \bibinfo {pages} {041103} (\bibinfo {year} {2019})},\
  \Eprint {http://arxiv.org/abs/1712.01396} {arXiv:1712.01396 [astro-ph.CO]}
  \BibitemShut {NoStop}%
\bibitem [{\citenamefont {Fan}\ \emph {et~al.}(2017)\citenamefont {Fan},
  \citenamefont {Liao}, \citenamefont {Biesiada}, \citenamefont
  {Piorkowska-Kurpas},\ and\ \citenamefont {Zhu}}]{Fan:2016swi}%
  \BibitemOpen
  \bibfield  {author} {\bibinfo {author} {\bibfnamefont {X.-L.}\ \bibnamefont
  {Fan}}, \bibinfo {author} {\bibfnamefont {K.}~\bibnamefont {Liao}}, \bibinfo
  {author} {\bibfnamefont {M.}~\bibnamefont {Biesiada}}, \bibinfo {author}
  {\bibfnamefont {A.}~\bibnamefont {Piorkowska-Kurpas}}, \ and\ \bibinfo
  {author} {\bibfnamefont {Z.-H.}\ \bibnamefont {Zhu}},\ }\href {\doibase
  10.1103/physrevlett.118.091102, 10.1103/PhysRevLett.118.091102} {\bibfield
  {journal} {\bibinfo  {journal} {Phys. Rev. Lett.}\ }\textbf {\bibinfo
  {volume} {118}},\ \bibinfo {pages} {091102} (\bibinfo {year} {2017})},\
  \Eprint {http://arxiv.org/abs/1612.04095} {arXiv:1612.04095 [gr-qc]}
  \BibitemShut {NoStop}%
\bibitem [{\citenamefont {Collett}\ and\ \citenamefont
  {Bacon}(2017)}]{Collett:2016dey}%
  \BibitemOpen
  \bibfield  {author} {\bibinfo {author} {\bibfnamefont {T.~E.}\ \bibnamefont
  {Collett}}\ and\ \bibinfo {author} {\bibfnamefont {D.}~\bibnamefont
  {Bacon}},\ }\href {\doibase 10.1103/PhysRevLett.118.091101} {\bibfield
  {journal} {\bibinfo  {journal} {Phys. Rev. Lett.}\ }\textbf {\bibinfo
  {volume} {118}},\ \bibinfo {pages} {091101} (\bibinfo {year} {2017})},\
  \Eprint {http://arxiv.org/abs/1602.05882} {arXiv:1602.05882 [astro-ph.HE]}
  \BibitemShut {NoStop}%
\bibitem [{\citenamefont {Sereno}\ \emph {et~al.}(2010)\citenamefont {Sereno},
  \citenamefont {Sesana}, \citenamefont {Bleuler}, \citenamefont {Jetzer},
  \citenamefont {Volonteri},\ and\ \citenamefont {Begelman}}]{Sereno:2010dr}%
  \BibitemOpen
  \bibfield  {author} {\bibinfo {author} {\bibfnamefont {M.}~\bibnamefont
  {Sereno}}, \bibinfo {author} {\bibfnamefont {A.}~\bibnamefont {Sesana}},
  \bibinfo {author} {\bibfnamefont {A.}~\bibnamefont {Bleuler}}, \bibinfo
  {author} {\bibfnamefont {P.}~\bibnamefont {Jetzer}}, \bibinfo {author}
  {\bibfnamefont {M.}~\bibnamefont {Volonteri}}, \ and\ \bibinfo {author}
  {\bibfnamefont {M.~C.}\ \bibnamefont {Begelman}},\ }\href {\doibase
  10.1103/PhysRevLett.105.251101} {\bibfield  {journal} {\bibinfo  {journal}
  {Phys. Rev. Lett.}\ }\textbf {\bibinfo {volume} {105}},\ \bibinfo {pages}
  {251101} (\bibinfo {year} {2010})},\ \Eprint {http://arxiv.org/abs/1011.5238}
  {arXiv:1011.5238 [astro-ph.CO]} \BibitemShut {NoStop}%
\bibitem [{\citenamefont {Sereno}\ \emph {et~al.}(2011)\citenamefont {Sereno},
  \citenamefont {Jetzer}, \citenamefont {Sesana},\ and\ \citenamefont
  {Volonteri}}]{Sereno:2011ty}%
  \BibitemOpen
  \bibfield  {author} {\bibinfo {author} {\bibfnamefont {M.}~\bibnamefont
  {Sereno}}, \bibinfo {author} {\bibfnamefont {P.}~\bibnamefont {Jetzer}},
  \bibinfo {author} {\bibfnamefont {A.}~\bibnamefont {Sesana}}, \ and\ \bibinfo
  {author} {\bibfnamefont {M.}~\bibnamefont {Volonteri}},\ }\href {\doibase
  10.1111/j.1365-2966.2011.18895.x} {\bibfield  {journal} {\bibinfo  {journal}
  {Mon. Not. Roy. Astron. Soc.}\ }\textbf {\bibinfo {volume} {415}},\ \bibinfo
  {pages} {2773} (\bibinfo {year} {2011})},\ \Eprint
  {http://arxiv.org/abs/1104.1977} {arXiv:1104.1977 [astro-ph.CO]} \BibitemShut
  {NoStop}%
\bibitem [{\citenamefont {Liao}\ \emph {et~al.}(2017)\citenamefont {Liao},
  \citenamefont {Fan}, \citenamefont {Ding}, \citenamefont {Biesiada},\ and\
  \citenamefont {Zhu}}]{Liao:2017ioi}%
  \BibitemOpen
  \bibfield  {author} {\bibinfo {author} {\bibfnamefont {K.}~\bibnamefont
  {Liao}}, \bibinfo {author} {\bibfnamefont {X.-L.}\ \bibnamefont {Fan}},
  \bibinfo {author} {\bibfnamefont {X.-H.}\ \bibnamefont {Ding}}, \bibinfo
  {author} {\bibfnamefont {M.}~\bibnamefont {Biesiada}}, \ and\ \bibinfo
  {author} {\bibfnamefont {Z.-H.}\ \bibnamefont {Zhu}},\ }\href {\doibase
  10.1038/s41467-017-01152-9, 10.1038/s41467-017-02135-6} {\bibfield  {journal}
  {\bibinfo  {journal} {Nature Commun.}\ }\textbf {\bibinfo {volume} {8}},\
  \bibinfo {pages} {1148} (\bibinfo {year} {2017})},\ \bibinfo {note}
  {[Erratum: Nature Commun.8,no.1,2136(2017)]},\ \Eprint
  {http://arxiv.org/abs/1703.04151} {arXiv:1703.04151 [astro-ph.CO]}
  \BibitemShut {NoStop}%
\bibitem [{\citenamefont {Dai}\ \emph {et~al.}(2018)\citenamefont {Dai},
  \citenamefont {Li}, \citenamefont {Zackay}, \citenamefont {Mao},\ and\
  \citenamefont {Lu}}]{Dai:2018enj}%
  \BibitemOpen
  \bibfield  {author} {\bibinfo {author} {\bibfnamefont {L.}~\bibnamefont
  {Dai}}, \bibinfo {author} {\bibfnamefont {S.-S.}\ \bibnamefont {Li}},
  \bibinfo {author} {\bibfnamefont {B.}~\bibnamefont {Zackay}}, \bibinfo
  {author} {\bibfnamefont {S.}~\bibnamefont {Mao}}, \ and\ \bibinfo {author}
  {\bibfnamefont {Y.}~\bibnamefont {Lu}},\ }\href {\doibase
  10.1103/PhysRevD.98.104029} {\bibfield  {journal} {\bibinfo  {journal} {Phys.
  Rev. D}\ }\textbf {\bibinfo {volume} {98}},\ \bibinfo {pages} {104029}
  (\bibinfo {year} {2018})},\ \Eprint {http://arxiv.org/abs/1810.00003}
  {arXiv:1810.00003 [gr-qc]} \BibitemShut {NoStop}%
\bibitem [{\citenamefont {Sun}\ and\ \citenamefont {Fan}(2019)}]{Sun:2019ztn}%
  \BibitemOpen
  \bibfield  {author} {\bibinfo {author} {\bibfnamefont {D.}~\bibnamefont
  {Sun}}\ and\ \bibinfo {author} {\bibfnamefont {X.}~\bibnamefont {Fan}},\
  }\href@noop {} {\  (\bibinfo {year} {2019})},\ \Eprint
  {http://arxiv.org/abs/1911.08268} {arXiv:1911.08268 [gr-qc]} \BibitemShut
  {NoStop}%
\bibitem [{\citenamefont {Hannuksela}\ \emph {et~al.}(2019)\citenamefont
  {Hannuksela}, \citenamefont {Haris}, \citenamefont {Ng}, \citenamefont
  {Kumar}, \citenamefont {Mehta}, \citenamefont {Keitel}, \citenamefont {Li},\
  and\ \citenamefont {Ajith}}]{Hannuksela:2019kle}%
  \BibitemOpen
  \bibfield  {author} {\bibinfo {author} {\bibfnamefont {O.~A.}\ \bibnamefont
  {Hannuksela}}, \bibinfo {author} {\bibfnamefont {K.}~\bibnamefont {Haris}},
  \bibinfo {author} {\bibfnamefont {K.~K.~Y.}\ \bibnamefont {Ng}}, \bibinfo
  {author} {\bibfnamefont {S.}~\bibnamefont {Kumar}}, \bibinfo {author}
  {\bibfnamefont {A.~K.}\ \bibnamefont {Mehta}}, \bibinfo {author}
  {\bibfnamefont {D.}~\bibnamefont {Keitel}}, \bibinfo {author} {\bibfnamefont
  {T.~G.~F.}\ \bibnamefont {Li}}, \ and\ \bibinfo {author} {\bibfnamefont
  {P.}~\bibnamefont {Ajith}},\ }\href {\doibase 10.3847/2041-8213/ab0c0f}
  {\bibfield  {journal} {\bibinfo  {journal} {Astrophys. J.}\ }\textbf
  {\bibinfo {volume} {874}},\ \bibinfo {pages} {L2} (\bibinfo {year} {2019})},\
  \Eprint {http://arxiv.org/abs/1901.02674} {arXiv:1901.02674 [gr-qc]}
  \BibitemShut {NoStop}%
\bibitem [{\citenamefont {Piran}\ and\ \citenamefont
  {Safier}(1985)}]{Piran1985nf}%
  \BibitemOpen
  \bibfield  {author} {\bibinfo {author} {\bibfnamefont {T.}~\bibnamefont
  {Piran}}\ and\ \bibinfo {author} {\bibfnamefont {P.~N.}\ \bibnamefont
  {Safier}},\ }\href {\doibase 10.1038/318271a0} {\bibfield  {journal}
  {\bibinfo  {journal} {Nature}\ }\textbf {\bibinfo {volume} {318}},\ \bibinfo
  {pages} {271} (\bibinfo {year} {1985})}\BibitemShut {NoStop}%
\bibitem [{\citenamefont {Piran}\ \emph {et~al.}(1985)\citenamefont {Piran},
  \citenamefont {Safier},\ and\ \citenamefont {Stark}}]{Piran:1985dk}%
  \BibitemOpen
  \bibfield  {author} {\bibinfo {author} {\bibfnamefont {T.}~\bibnamefont
  {Piran}}, \bibinfo {author} {\bibfnamefont {P.~N.}\ \bibnamefont {Safier}}, \
  and\ \bibinfo {author} {\bibfnamefont {R.~F.}\ \bibnamefont {Stark}},\ }\href
  {\doibase 10.1103/PhysRevD.32.3101} {\bibfield  {journal} {\bibinfo
  {journal} {Phys. Rev. D}\ }\textbf {\bibinfo {volume} {32}},\ \bibinfo
  {pages} {3101} (\bibinfo {year} {1985})}\BibitemShut {NoStop}%
\bibitem [{\citenamefont {Wang}(1991)}]{Wang:1991nf}%
  \BibitemOpen
  \bibfield  {author} {\bibinfo {author} {\bibfnamefont {A.}~\bibnamefont
  {Wang}},\ }\href {\doibase 10.1103/PhysRevD.44.1120} {\bibfield  {journal}
  {\bibinfo  {journal} {Phys. Rev. D}\ }\textbf {\bibinfo {volume} {44}},\
  \bibinfo {pages} {1120} (\bibinfo {year} {1991})}\BibitemShut {NoStop}%
\bibitem [{\citenamefont {Yang}\ \emph {et~al.}(2019)\citenamefont {Yang},
  \citenamefont {Ding}, \citenamefont {Biesiada}, \citenamefont {Liao},\ and\
  \citenamefont {Zhu}}]{Yang:2019jhw}%
  \BibitemOpen
  \bibfield  {author} {\bibinfo {author} {\bibfnamefont {L.}~\bibnamefont
  {Yang}}, \bibinfo {author} {\bibfnamefont {X.}~\bibnamefont {Ding}}, \bibinfo
  {author} {\bibfnamefont {M.}~\bibnamefont {Biesiada}}, \bibinfo {author}
  {\bibfnamefont {K.}~\bibnamefont {Liao}}, \ and\ \bibinfo {author}
  {\bibfnamefont {Z.-H.}\ \bibnamefont {Zhu}},\ }\href {\doibase
  10.3847/1538-4357/ab095c} {\bibfield  {journal} {\bibinfo  {journal}
  {Astrophys. J.}\ }\textbf {\bibinfo {volume} {874}},\ \bibinfo {pages} {139}
  (\bibinfo {year} {2019})},\ \Eprint {http://arxiv.org/abs/1903.11079}
  {arXiv:1903.11079 [astro-ph.GA]} \BibitemShut {NoStop}%
\bibitem [{\citenamefont {Turner}\ \emph {et~al.}(1984)\citenamefont {Turner},
  \citenamefont {Ostriker},\ and\ \citenamefont {Gott}}]{Turner:1984ch}%
  \BibitemOpen
  \bibfield  {author} {\bibinfo {author} {\bibfnamefont {E.~L.}\ \bibnamefont
  {Turner}}, \bibinfo {author} {\bibfnamefont {J.~P.}\ \bibnamefont
  {Ostriker}}, \ and\ \bibinfo {author} {\bibfnamefont {J.~R.}\ \bibnamefont
  {Gott}, \bibfnamefont {III}},\ }\href {\doibase 10.1086/162379} {\bibfield
  {journal} {\bibinfo  {journal} {Astrophys. J.}\ }\textbf {\bibinfo {volume}
  {284}},\ \bibinfo {pages} {1} (\bibinfo {year} {1984})}\BibitemShut {NoStop}%
\bibitem [{\citenamefont {Moeller}\ \emph {et~al.}(2007)\citenamefont
  {Moeller}, \citenamefont {Kitzbichler},\ and\ \citenamefont
  {Natarajan}}]{Moeller:2006cu}%
  \BibitemOpen
  \bibfield  {author} {\bibinfo {author} {\bibfnamefont {O.}~\bibnamefont
  {Moeller}}, \bibinfo {author} {\bibfnamefont {M.}~\bibnamefont
  {Kitzbichler}}, \ and\ \bibinfo {author} {\bibfnamefont {P.}~\bibnamefont
  {Natarajan}},\ }\href {\doibase 10.1111/j.1365-2966.2007.12004.x} {\bibfield
  {journal} {\bibinfo  {journal} {Mon. Not. Roy. Astron. Soc.}\ }\textbf
  {\bibinfo {volume} {379}},\ \bibinfo {pages} {1195} (\bibinfo {year}
  {2007})},\ \Eprint {http://arxiv.org/abs/astro-ph/0607032}
  {arXiv:astro-ph/0607032 [astro-ph]} \BibitemShut {NoStop}%
\bibitem [{\citenamefont {Oguri}(2018)}]{Oguri:2018muv}%
  \BibitemOpen
  \bibfield  {author} {\bibinfo {author} {\bibfnamefont {M.}~\bibnamefont
  {Oguri}},\ }\href {\doibase 10.1093/mnras/sty2145} {\bibfield  {journal}
  {\bibinfo  {journal} {Mon. Not. Roy. Astron. Soc.}\ }\textbf {\bibinfo
  {volume} {480}},\ \bibinfo {pages} {3842} (\bibinfo {year} {2018})},\ \Eprint
  {http://arxiv.org/abs/1807.02584} {arXiv:1807.02584 [astro-ph.CO]}
  \BibitemShut {NoStop}%
\bibitem [{\citenamefont {Isi}\ and\ \citenamefont
  {Weinstein}(2017)}]{Isi:2017fbj}%
  \BibitemOpen
  \bibfield  {author} {\bibinfo {author} {\bibfnamefont {M.}~\bibnamefont
  {Isi}}\ and\ \bibinfo {author} {\bibfnamefont {A.~J.}\ \bibnamefont
  {Weinstein}},\ }\href@noop {} {\  (\bibinfo {year} {2017})},\ \Eprint
  {http://arxiv.org/abs/1710.03794} {arXiv:1710.03794 [gr-qc]} \BibitemShut
  {NoStop}%
\bibitem [{\citenamefont {Maggiore}(2007)}]{Maggiore:1900zz}%
  \BibitemOpen
  \bibfield  {author} {\bibinfo {author} {\bibfnamefont {M.}~\bibnamefont
  {Maggiore}},\ }\href {http://www.oup.com/uk/catalogue/?ci=9780198570745}
  {\emph {\bibinfo {title} {{Gravitational Waves. Vol. 1: Theory and
  Experiments}}}},\ Oxford Master Series in Physics\ (\bibinfo  {publisher}
  {Oxford University Press},\ \bibinfo {year} {2007})\BibitemShut {NoStop}%
\bibitem [{\citenamefont {Dal~Canton}\ \emph {et~al.}(2014)\citenamefont
  {Dal~Canton} \emph {et~al.}}]{Canton:2014ena}%
  \BibitemOpen
  \bibfield  {author} {\bibinfo {author} {\bibfnamefont {T.}~\bibnamefont
  {Dal~Canton}} \emph {et~al.},\ }\href {\doibase 10.1103/PhysRevD.90.082004}
  {\bibfield  {journal} {\bibinfo  {journal} {Phys. Rev. D}\ }\textbf {\bibinfo
  {volume} {90}},\ \bibinfo {pages} {082004} (\bibinfo {year} {2014})},\
  \Eprint {http://arxiv.org/abs/1405.6731} {arXiv:1405.6731 [gr-qc]}
  \BibitemShut {NoStop}%
\bibitem [{\citenamefont {Usman}\ \emph {et~al.}(2016)\citenamefont {Usman}
  \emph {et~al.}}]{Usman:2015kfa}%
  \BibitemOpen
  \bibfield  {author} {\bibinfo {author} {\bibfnamefont {S.~A.}\ \bibnamefont
  {Usman}} \emph {et~al.},\ }\href {\doibase 10.1088/0264-9381/33/21/215004}
  {\bibfield  {journal} {\bibinfo  {journal} {Class. Quant. Grav.}\ }\textbf
  {\bibinfo {volume} {33}},\ \bibinfo {pages} {215004} (\bibinfo {year}
  {2016})},\ \Eprint {http://arxiv.org/abs/1508.02357} {arXiv:1508.02357
  [gr-qc]} \BibitemShut {NoStop}%
\bibitem [{\citenamefont {Nitz}\ \emph {et~al.}(2019)\citenamefont {Nitz} \emph
  {et~al.}}]{alex_nitz_2019_2643618s}%
  \BibitemOpen
  \bibfield  {author} {\bibinfo {author} {\bibfnamefont {A.}~\bibnamefont
  {Nitz}} \emph {et~al.},\ }\href {\doibase 10.5281/zenodo.2643618} {\enquote
  {\bibinfo {title} {gwastro/pycbc: Pycbc release v1.13.6},}\ } (\bibinfo
  {year} {2019})\BibitemShut {NoStop}%
\bibitem [{\citenamefont {Yunes}\ and\ \citenamefont
  {Siemens}(2013)}]{Yunes:2013dva}%
  \BibitemOpen
  \bibfield  {author} {\bibinfo {author} {\bibfnamefont {N.}~\bibnamefont
  {Yunes}}\ and\ \bibinfo {author} {\bibfnamefont {X.}~\bibnamefont
  {Siemens}},\ }\href {\doibase 10.12942/lrr-2013-9} {\bibfield  {journal}
  {\bibinfo  {journal} {Living Rev. Rel.}\ }\textbf {\bibinfo {volume} {16}},\
  \bibinfo {pages} {9} (\bibinfo {year} {2013})},\ \Eprint
  {http://arxiv.org/abs/1304.3473} {arXiv:1304.3473 [gr-qc]} \BibitemShut
  {NoStop}%
\bibitem [{\citenamefont {Punturo}\ \emph {et~al.}(2010)\citenamefont {Punturo}
  \emph {et~al.}}]{Punturo:2010zza}%
  \BibitemOpen
  \bibfield  {author} {\bibinfo {author} {\bibfnamefont {M.}~\bibnamefont
  {Punturo}} \emph {et~al.},\ }\bibfield  {booktitle} {\emph {\bibinfo
  {booktitle} {{Gravitational waves. Proceedings, 8th Edoardo Amaldi
  Conference, Amaldi 8, New York, USA, June 22-26, 2009}}},\ }\href {\doibase
  10.1088/0264-9381/27/8/084007} {\bibfield  {journal} {\bibinfo  {journal}
  {Class. Quant. Grav.}\ }\textbf {\bibinfo {volume} {27}},\ \bibinfo {pages}
  {084007} (\bibinfo {year} {2010})}\BibitemShut {NoStop}%
\bibitem [{\citenamefont {Abbott}\ \emph
  {et~al.}(2017{\natexlab{e}})\citenamefont {Abbott} \emph
  {et~al.}}]{Evans:2016mbw}%
  \BibitemOpen
  \bibfield  {author} {\bibinfo {author} {\bibfnamefont {B.~P.}\ \bibnamefont
  {Abbott}} \emph {et~al.} (\bibinfo {collaboration} {LIGO Scientific}),\
  }\href {\doibase 10.1088/1361-6382/aa51f4} {\bibfield  {journal} {\bibinfo
  {journal} {Class. Quant. Grav.}\ }\textbf {\bibinfo {volume} {34}},\ \bibinfo
  {pages} {044001} (\bibinfo {year} {2017}{\natexlab{e}})},\ \Eprint
  {http://arxiv.org/abs/1607.08697} {arXiv:1607.08697 [astro-ph.IM]}
  \BibitemShut {NoStop}%
\bibitem [{\citenamefont {Mirshekari}\ \emph {et~al.}(2012)\citenamefont
  {Mirshekari}, \citenamefont {Yunes},\ and\ \citenamefont
  {Will}}]{Mirshekari:2011yq}%
  \BibitemOpen
  \bibfield  {author} {\bibinfo {author} {\bibfnamefont {S.}~\bibnamefont
  {Mirshekari}}, \bibinfo {author} {\bibfnamefont {N.}~\bibnamefont {Yunes}}, \
  and\ \bibinfo {author} {\bibfnamefont {C.~M.}\ \bibnamefont {Will}},\ }\href
  {\doibase 10.1103/PhysRevD.85.024041} {\bibfield  {journal} {\bibinfo
  {journal} {Phys. Rev. D}\ }\textbf {\bibinfo {volume} {85}},\ \bibinfo
  {pages} {024041} (\bibinfo {year} {2012})},\ \Eprint
  {http://arxiv.org/abs/1110.2720} {arXiv:1110.2720 [gr-qc]} \BibitemShut
  {NoStop}%
\bibitem [{\citenamefont {Bonvin}\ \emph {et~al.}(2017)\citenamefont {Bonvin},
  \citenamefont {Caprini}, \citenamefont {Sturani},\ and\ \citenamefont
  {Tamanini}}]{Bonvin:2016qxr}%
  \BibitemOpen
  \bibfield  {author} {\bibinfo {author} {\bibfnamefont {C.}~\bibnamefont
  {Bonvin}}, \bibinfo {author} {\bibfnamefont {C.}~\bibnamefont {Caprini}},
  \bibinfo {author} {\bibfnamefont {R.}~\bibnamefont {Sturani}}, \ and\
  \bibinfo {author} {\bibfnamefont {N.}~\bibnamefont {Tamanini}},\ }\href
  {\doibase 10.1103/PhysRevD.95.044029} {\bibfield  {journal} {\bibinfo
  {journal} {Phys. Rev. D}\ }\textbf {\bibinfo {volume} {95}},\ \bibinfo
  {pages} {044029} (\bibinfo {year} {2017})},\ \Eprint
  {http://arxiv.org/abs/1609.08093} {arXiv:1609.08093 [astro-ph.CO]}
  \BibitemShut {NoStop}%
\bibitem [{\citenamefont {Ding}\ \emph {et~al.}(2015)\citenamefont {Ding},
  \citenamefont {Biesiada},\ and\ \citenamefont {Zhu}}]{Ding:2015uha}%
  \BibitemOpen
  \bibfield  {author} {\bibinfo {author} {\bibfnamefont {X.}~\bibnamefont
  {Ding}}, \bibinfo {author} {\bibfnamefont {M.}~\bibnamefont {Biesiada}}, \
  and\ \bibinfo {author} {\bibfnamefont {Z.-H.}\ \bibnamefont {Zhu}},\ }\href
  {\doibase 10.1088/1475-7516/2015/12/006} {\bibfield  {journal} {\bibinfo
  {journal} {JCAP}\ }\textbf {\bibinfo {volume} {1512}},\ \bibinfo {pages}
  {006} (\bibinfo {year} {2015})},\ \Eprint {http://arxiv.org/abs/1508.05000}
  {arXiv:1508.05000 [astro-ph.HE]} \BibitemShut {NoStop}%
\bibitem [{\citenamefont {Li}\ \emph {et~al.}(2018)\citenamefont {Li},
  \citenamefont {Mao}, \citenamefont {Zhao},\ and\ \citenamefont
  {Lu}}]{Li:2018prc}%
  \BibitemOpen
  \bibfield  {author} {\bibinfo {author} {\bibfnamefont {S.-S.}\ \bibnamefont
  {Li}}, \bibinfo {author} {\bibfnamefont {S.}~\bibnamefont {Mao}}, \bibinfo
  {author} {\bibfnamefont {Y.}~\bibnamefont {Zhao}}, \ and\ \bibinfo {author}
  {\bibfnamefont {Y.}~\bibnamefont {Lu}},\ }\href {\doibase
  10.1093/mnras/sty411} {\bibfield  {journal} {\bibinfo  {journal} {Mon. Not.
  Roy. Astron. Soc.}\ }\textbf {\bibinfo {volume} {476}},\ \bibinfo {pages}
  {2220} (\bibinfo {year} {2018})},\ \Eprint {http://arxiv.org/abs/1802.05089}
  {arXiv:1802.05089 [astro-ph.CO]} \BibitemShut {NoStop}%
\bibitem [{\citenamefont {Pi\'orkowska}\ \emph {et~al.}(2013)\citenamefont
  {Pi\'orkowska}, \citenamefont {Biesiada},\ and\ \citenamefont
  {Zhu}}]{Piorkowska:2013eww}%
  \BibitemOpen
  \bibfield  {author} {\bibinfo {author} {\bibfnamefont {A.}~\bibnamefont
  {Pi\'orkowska}}, \bibinfo {author} {\bibfnamefont {M.}~\bibnamefont
  {Biesiada}}, \ and\ \bibinfo {author} {\bibfnamefont {Z.-H.}\ \bibnamefont
  {Zhu}},\ }\href {\doibase 10.1088/1475-7516/2013/10/022} {\bibfield
  {journal} {\bibinfo  {journal} {JCAP}\ }\textbf {\bibinfo {volume} {1310}},\
  \bibinfo {pages} {022} (\bibinfo {year} {2013})},\ \Eprint
  {http://arxiv.org/abs/1309.5731} {arXiv:1309.5731 [astro-ph.CO]} \BibitemShut
  {NoStop}%
\bibitem [{\citenamefont {Gong}(2019)}]{gong_yungui_2019_2574620}%
  \BibitemOpen
  \bibfield  {author} {\bibinfo {author} {\bibfnamefont {Y.}~\bibnamefont
  {Gong}},\ }\href {\doibase 10.5281/zenodo.2574620} {\enquote {\bibinfo
  {title} {{The supplementary material for antenna transfer function (arXiv:
  1901.09624)}},}\ } (\bibinfo {year} {2019})\BibitemShut {NoStop}%
\bibitem [{\citenamefont {Robson}\ \emph {et~al.}(2019)\citenamefont {Robson},
  \citenamefont {Cornish},\ and\ \citenamefont {Liug}}]{Cornish:2018dyw}%
  \BibitemOpen
  \bibfield  {author} {\bibinfo {author} {\bibfnamefont {T.}~\bibnamefont
  {Robson}}, \bibinfo {author} {\bibfnamefont {N.~J.}\ \bibnamefont {Cornish}},
  \ and\ \bibinfo {author} {\bibfnamefont {C.}~\bibnamefont {Liug}},\ }\href
  {\doibase 10.1088/1361-6382/ab1101} {\bibfield  {journal} {\bibinfo
  {journal} {Class. Quant. Grav.}\ }\textbf {\bibinfo {volume} {36}},\ \bibinfo
  {pages} {105011} (\bibinfo {year} {2019})},\ \Eprint
  {http://arxiv.org/abs/1803.01944} {arXiv:1803.01944 [astro-ph.HE]}
  \BibitemShut {NoStop}%
\bibitem [{\citenamefont {Cornish}(2019)}]{neil_cornish_2019_2636514}%
  \BibitemOpen
  \bibfield  {author} {\bibinfo {author} {\bibfnamefont {N.}~\bibnamefont
  {Cornish}},\ }\href {\doibase 10.5281/zenodo.2636514} {\enquote {\bibinfo
  {title} {{eXtremeGravityInstitute/LISA Sensitivity: Version 1}},}\ }
  (\bibinfo {year} {2019})\BibitemShut {NoStop}%
\bibitem [{\citenamefont {Weinberg}(2008)}]{Weinberg:2008zzc}%
  \BibitemOpen
  \bibfield  {author} {\bibinfo {author} {\bibfnamefont {S.}~\bibnamefont
  {Weinberg}},\ }\href {http://www.oup.com/uk/catalogue/?ci=9780198526827}
  {\emph {\bibinfo {title} {{Cosmology}}}}\ (\bibinfo  {publisher} {Oxford, UK:
  Oxford Univ. Pr. (2008) 593 p},\ \bibinfo {year} {2008})\BibitemShut
  {NoStop}%
\bibitem [{\citenamefont {Liu}\ \emph {et~al.}(2019{\natexlab{b}})\citenamefont
  {Liu}, \citenamefont {Li},\ and\ \citenamefont {Zhu}}]{Liu:2019dds}%
  \BibitemOpen
  \bibfield  {author} {\bibinfo {author} {\bibfnamefont {B.}~\bibnamefont
  {Liu}}, \bibinfo {author} {\bibfnamefont {Z.}~\bibnamefont {Li}}, \ and\
  \bibinfo {author} {\bibfnamefont {Z.-H.}\ \bibnamefont {Zhu}},\ }\href
  {\doibase 10.1093/mnras/stz1179} {\bibfield  {journal} {\bibinfo  {journal}
  {Mon. Not. Roy. Astron. Soc.}\ }\textbf {\bibinfo {volume} {487}},\ \bibinfo
  {pages} {1980} (\bibinfo {year} {2019}{\natexlab{b}})},\ \Eprint
  {http://arxiv.org/abs/1904.11751} {arXiv:1904.11751 [astro-ph.CO]}
  \BibitemShut {NoStop}%
\bibitem [{\citenamefont {Sesana}\ \emph {et~al.}(2009)\citenamefont {Sesana},
  \citenamefont {Volonteri},\ and\ \citenamefont {Haardt}}]{Sesana:2008ur}%
  \BibitemOpen
  \bibfield  {author} {\bibinfo {author} {\bibfnamefont {A.}~\bibnamefont
  {Sesana}}, \bibinfo {author} {\bibfnamefont {M.}~\bibnamefont {Volonteri}}, \
  and\ \bibinfo {author} {\bibfnamefont {F.}~\bibnamefont {Haardt}},\
  }\bibfield  {booktitle} {\emph {\bibinfo {booktitle} {{Laser Interferometer
  Space Antenna. Proceedings, 7th international LISA Symposium, Barcelona,
  Spain, June 16-20, 2008}}},\ }\href {\doibase 10.1088/0264-9381/26/9/094033}
  {\bibfield  {journal} {\bibinfo  {journal} {Class. Quant. Grav.}\ }\textbf
  {\bibinfo {volume} {26}},\ \bibinfo {pages} {094033} (\bibinfo {year}
  {2009})},\ \Eprint {http://arxiv.org/abs/0810.5554} {arXiv:0810.5554
  [astro-ph]} \BibitemShut {NoStop}%
\bibitem [{\citenamefont {Klein}\ \emph {et~al.}(2016)\citenamefont {Klein}
  \emph {et~al.}}]{Klein:2015hvg}%
  \BibitemOpen
  \bibfield  {author} {\bibinfo {author} {\bibfnamefont {A.}~\bibnamefont
  {Klein}} \emph {et~al.},\ }\href {\doibase 10.1103/PhysRevD.93.024003}
  {\bibfield  {journal} {\bibinfo  {journal} {Phys. Rev. D}\ }\textbf {\bibinfo
  {volume} {93}},\ \bibinfo {pages} {024003} (\bibinfo {year} {2016})},\
  \Eprint {http://arxiv.org/abs/1511.05581} {arXiv:1511.05581 [gr-qc]}
  \BibitemShut {NoStop}%
\bibitem [{\citenamefont {Gerosa}\ \emph {et~al.}(2019)\citenamefont {Gerosa},
  \citenamefont {Ma}, \citenamefont {Wong}, \citenamefont {Berti},
  \citenamefont {O'Shaughnessy}, \citenamefont {Chen},\ and\ \citenamefont
  {Belczynski}}]{Gerosa:2019dbe}%
  \BibitemOpen
  \bibfield  {author} {\bibinfo {author} {\bibfnamefont {D.}~\bibnamefont
  {Gerosa}}, \bibinfo {author} {\bibfnamefont {S.}~\bibnamefont {Ma}}, \bibinfo
  {author} {\bibfnamefont {K.~W.~K.}\ \bibnamefont {Wong}}, \bibinfo {author}
  {\bibfnamefont {E.}~\bibnamefont {Berti}}, \bibinfo {author} {\bibfnamefont
  {R.}~\bibnamefont {O'Shaughnessy}}, \bibinfo {author} {\bibfnamefont
  {Y.}~\bibnamefont {Chen}}, \ and\ \bibinfo {author} {\bibfnamefont
  {K.}~\bibnamefont {Belczynski}},\ }\href {\doibase
  10.1103/PhysRevD.99.103004} {\bibfield  {journal} {\bibinfo  {journal} {Phys.
  Rev. D}\ }\textbf {\bibinfo {volume} {99}},\ \bibinfo {pages} {103004}
  (\bibinfo {year} {2019})},\ \Eprint {http://arxiv.org/abs/1902.00021}
  {arXiv:1902.00021 [astro-ph.HE]} \BibitemShut {NoStop}%
\bibitem [{\citenamefont {Boco}\ \emph {et~al.}(2019)\citenamefont {Boco},
  \citenamefont {Lapi}, \citenamefont {Goswami}, \citenamefont {Perrotta},
  \citenamefont {Baccigalupi},\ and\ \citenamefont {Danese}}]{Boco_2019}%
  \BibitemOpen
  \bibfield  {author} {\bibinfo {author} {\bibfnamefont {L.}~\bibnamefont
  {Boco}}, \bibinfo {author} {\bibfnamefont {A.}~\bibnamefont {Lapi}}, \bibinfo
  {author} {\bibfnamefont {S.}~\bibnamefont {Goswami}}, \bibinfo {author}
  {\bibfnamefont {F.}~\bibnamefont {Perrotta}}, \bibinfo {author}
  {\bibfnamefont {C.}~\bibnamefont {Baccigalupi}}, \ and\ \bibinfo {author}
  {\bibfnamefont {L.}~\bibnamefont {Danese}},\ }\href {\doibase
  10.3847/1538-4357/ab328e} {\bibfield  {journal} {\bibinfo  {journal}
  {Astrophys. J.}\ }\textbf {\bibinfo {volume} {881}},\ \bibinfo {pages} {157}
  (\bibinfo {year} {2019})},\ \Eprint {http://arxiv.org/abs/1907.06841}
  {arXiv:1907.06841 [astro-ph.GA]} \BibitemShut {NoStop}%
\bibitem [{\citenamefont {Diego}\ \emph {et~al.}(2018)\citenamefont {Diego}
  \emph {et~al.}}]{Diego:2017drh}%
  \BibitemOpen
  \bibfield  {author} {\bibinfo {author} {\bibfnamefont {J.~M.}\ \bibnamefont
  {Diego}} \emph {et~al.},\ }\href {\doibase 10.3847/1538-4357/aab617}
  {\bibfield  {journal} {\bibinfo  {journal} {Astrophys. J.}\ }\textbf
  {\bibinfo {volume} {857}},\ \bibinfo {pages} {25} (\bibinfo {year} {2018})},\
  \Eprint {http://arxiv.org/abs/1706.10281} {arXiv:1706.10281 [astro-ph.CO]}
  \BibitemShut {NoStop}%
\end{thebibliography}%

\end{document}